\documentclass{aa}

\usepackage{verbatim}
\usepackage{graphicx}
\usepackage{epstopdf}
\usepackage{subfigure}
\usepackage{amsmath} 
\usepackage{color,soul}

\newcommand{\mdot}{\mbox{M$_\odot$~yr$^{-1}$}}

\newcommand{\kms}{\mbox{km~s$^{-1}$}}
\newcommand{\rsun}{\mbox{R$_{\odot}$}}

\title{Colliding Winds in Low-Mass Binary Star Systems: wind interactions and implications for habitable planets}

\titlerunning{Colliding Binary Winds and Habitability}

\author{C. P. Johnstone\inst{\ref{vienna}} \and A. Zhilkin\inst{\ref{russia1},\ref{russia2}} \and E. Pilat-Lohinger\inst{\ref{vienna}, \ref{graz}} \and D. Bisikalo\inst{\ref{russia1}} \and M. G\"{u}del\inst{\ref{vienna}} \and S. Eggl\inst{\ref{france}}}

\institute{University of Vienna, Department of Astrophysics, T\"{u}rkenschanzstrasse 17, 1180 Vienna, Austria \label{vienna}
\and Institute of Astronomy, Russian Academy of Sciences, Moscow, Russia \label{russia1}
\and Chelyabinsk State University, Chelyabinsk, Russia \label{russia2}
\and University of Graz, Institute of Physics, Universit\"{a}tsplatz 5, 8010 Graz, Austria \label{graz}
\and IMCCE Observatroire de Paris, UPMC, Universit\'{e} Lille 1, 75014 Paris, France \label{france}
}

\abstract
{
In binary star systems, the winds from the two components impact each other, leading to strong shocks and regions of enhanced density and temperature.
Potentially habitable circumbinary planets must continually be exposed to these interactions regions.
}
{
We study, for the first time, the interactions between winds from low-mass stars in a binary system, to show the wind conditions seen by potentially habitable circumbinary planets.
}
{ 
We use the advanced 3D numerical hydrodynamic code Nurgush to model the wind interactions of two identical winds from two solar mass stars with circular orbits and a binary separation of 0.5~AU.
As input into this model, we use a 1D hydrodynamic simulation of the solar wind, run using the Versatile Advection Code.  
We derive the locations of stable and habitable orbits in this system to explore what wind conditions potentially habitable planets will be exposed to during their orbits.
}
{
Our wind interaction simulations result in the formation of two strong shock waves separated by a region of enhanced density and temperature.
The wind-wind interaction region has a spiral shape due to Coriolis forces generated by the orbital motions of the two stars.
The stable and habitable zone in this system extends from $\sim$1.4~AU to $\sim$2.4~AU.
Habitable planets have to pass through strong shock waves several times per orbit and spend a significant amount of time embedded in the higher density matter between the shocks. 
The enhanced density in the wind-wind interaction region is likely to lead to a 20\% decrease in the size of a planet's magnetosphere.
}
{
Our results indicate that wind-wind interactions are likely to influence the magnetospheres and upper atmospheres of circumbinary planets and could have moderate implications for the development of habitable planetary environments. 
}

\begin{document}

\maketitle


\section{Introduction}

The formation and evolution of a habitable planetary atmosphere is highly sensitive to the surrounding stellar environment. 
The classical view of habitability is based on the ability for a planet to possess liquid water, which depends on the amount of visible light from the central star incident on the planet (\mbox{\citealt{1993Icar..101..108K}}; \mbox{\citealt{2013ApJ...765..131K}}).
Beyond visible light, stellar output also includes high energy radiation, such as EUV and \mbox{X-rays}, high energy particles, and stellar plasma in the form of coronal mass ejections and stellar winds.
All of this output can significantly influence the development of habitable planetary environments in ways that are currently poorly understood. 
For example, stellar \mbox{X-rays} and EUV can ionise and expand the planetary atmospheres and drive hydrodynamic mass-loss (\mbox{\citealt{1981Icar...48..150W}}; \mbox{\citealt{2003ApJ...598L.121L}}; \mbox{\citealt{2005ApJ...621.1049T}}; \mbox{\citealt{2008JGRE..113.5008T}}; \mbox{\citealt{2013MNRAS.430.1247L}}) and winds can compress the planetary magnetospheres and strip away atmospheric particles (\mbox{\citealt{2012ApJ...744...70K}}; \mbox{\citealt{2014A&A...562A.116K}}).

Since a large fraction of low-mass stars exist in multiple-star systems (\mbox{\citealt{1991A&A...248..485D}}; \mbox{\citealt{2006ApJ...640L..63L}}), an important question is the possibility of planetary habitability in binary systems.
This is a more complicated question due to the gravitational interactions between the planet and the two stars and the presence of output from both stars.
Planets in binary systems can have two types of orbital configurations: these are called \mbox{S-type} and \mbox{P-type}.
\mbox{S-type} orbits are circumstellar orbits where the planet orbits one of the stellar components of the binary system.
\mbox{P-type} orbits are circumbinary orbits where the planet orbits both of the stars.
Primarily as a result of the \emph{Kepler} mission, planets in both types of binary systems have been discovered (e.g. \mbox{\citealt{2011Sci...333.1602D}}; \mbox{\citealt{2012Natur.481..475W}}; \mbox{\citealt{2012Natur.491..207D}}; \mbox{\citealt{2014ApJ...784...14K}}). 
Although stable planetary orbits are possible in both types of systems, a necessary condition for habitability is that the regions of orbital stability coincide with the habitable zones. 
For example, \mbox{\citet{2013MNRAS.428.3104E}} studied dynamical and radiative star-planet interactions in 19 binary systems with different stellar masses and orbital parameters and found that in two of the systems, dynamically stable orbits are not possible within the habitable zones. 
An additional factor that can potentially affect the atmospheres of planets in \mbox{P-type} orbits is tidal interactions, which can influence the rotational evolution of both stars, and therefore their levels of magnetic activity.
Since high energy radiation can influence the evolution of a planetary atmosphere, this could be a significant factor in determining the formation of habitable environments in a subset of tight binary systems (\mbox{\citealt{2013ApJ...774L..26M}}).
A currently unexplored topic is the influences of winds, and specifically wind-wind interactions, on potentially habitable planets in binary systems.
This is especially interesting for \mbox{P-type} orbits since circumbinary planets will pass through wind-wind interaction regions every orbit.

As we discuss in detail in Section~\ref{sect:CWBintro}, colliding winds in binary star systems consisting of high-mass stars have been researched in great detail. 
These systems are known as colliding-wind binaries (CWBs), and typically contain combinations of Wolf-Rayet (\mbox{W-R}) and OB-type stars.  
The fundamental physical mechanisms driving the winds of high-mass stars differ significantly from those of low-mass stellar winds. 
Unlike the solar wind, the winds of massive stars are radiation driven, primarily through line absorption (\mbox{\citealt{1999isw..book.....L}}).
Such winds are called line-driven winds and can have speeds of several thousand~\kms and mass loss rates that exceed $10^{-5}$~\mdot, nine orders of magnitude higher than the mass loss rate of the current solar wind.
Extreme mass loss of this type can have significant effects on the evolution and internal structure of high-mass stars and can be easily observed (e.g. \mbox{\citealt{1996A&A...305..171P}}; \mbox{\citealt{2007A&A...473..603M}}).

The driving mechanisms responsible for the winds of low-mass stars are currently poorly understood. 
Such winds are likely to be analogous to the solar wind, which has a mass loss rate $\sim10^{-14}$~M$_\odot$~yr$^{-1}$ (\mbox{\citealt{2011MNRAS.417.2592C}}) and is known to be driven by the solar magnetic field.
Although other acceleration mechanisms, such as wave pressure, are probably required, the solar wind is thought to be accelerated primarily by thermal pressure gradients (\mbox{\citealt{1958ApJ...128..664P}}; \mbox{\citealt{2004ESASP.575..154C}}).
This requires that the wind is heated as it expands into and beyond the solar corona.
The main candidates for the heating of the wind are energy dissipation in the corona by Alfv\'{e}n waves generated by turbulent convective motion in the photosphere and energy dissipation by magnetic reconnection of stressed closed magnetic loops.
For a review of these mechanisms, see \mbox{\citet{2009LRSP....6....3C}}.
The resulting wind streams radially away from the Sun in all directions at all times with typical speeds of 400~\kms and 800~\kms. 
Despite the fact that the wind speed is non-isotropic and variable in time, the mass flux in the wind far from the solar surface is approximately isotropic and constant (\mbox{\citealt{1995ApJ...449L.157W}}; \mbox{\citealt{2010ApJ...715L.121W}}).

%
%

In this paper, we model for the first time the wind-wind interactions in a low-mass binary star system.
Our chosen system consists of two solar mass stars in a tight 0.5~AU orbit.
We link our results to calculations of the locations of stable planetary orbits and habitable zones to show what wind conditions a circumbinary planet in the habitable zone would see over the course of its orbit. 
In Section~\ref{sect:CWBintro}, we give an introduction to the subject of colliding-winds.  
In Section~\ref{sect:model}, we introduce our numerical models for wind formation and propagation, and for wind-wind interactions. 
In Section~\ref{sect:windresults}, we present the results of our wind model. 
In Section~\ref{sect:binaryhabitability}, we derive the locations of stable and habitable zones in our chosen system. 
Finally, in Section~\ref{sect:conclusions}, we discuss our results and the implications they could have for planetary habitability in tight binary systems.

\section{Colliding Wind Binaries: Observations and Theory} \label{sect:CWBintro}

\begin{figure}
\includegraphics[width=0.49\textwidth]{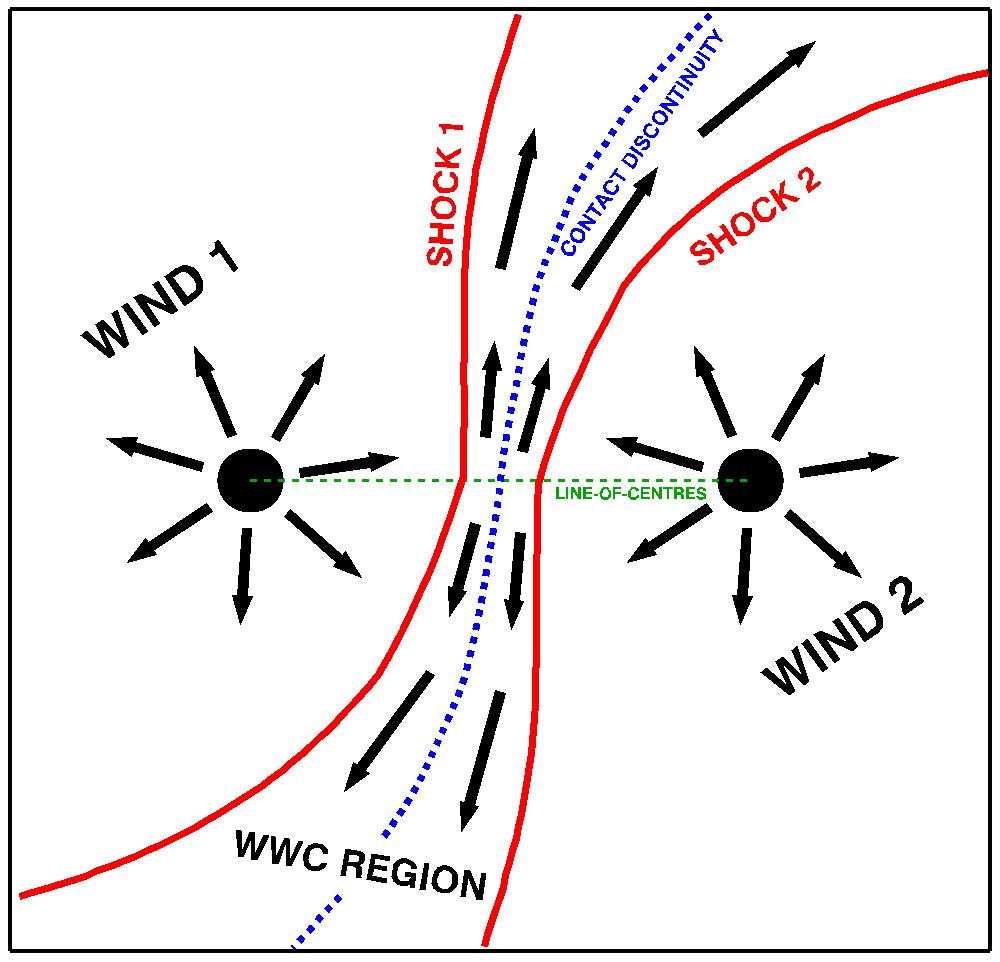}
\caption{
Cartoon showing the main features of wind-wind interactions in a binary system with two identical winds.
The wind-wind collision (WWC) region takes the form of two shock waves (\emph{red}) separated by a contact discontinuity (\emph{blue}) and has a spiral geometry due to the orbital motions of the two stars. 
Wind material flowing away from one of the stars at supersonic velocities impacts the closest shock and is decelerated and thermalised.
In the post shock gas, the material is then accelerated away from the centre of the WWC~region. 
}
 \label{fig:CWBcartoon}
\end{figure}


The colliding wind system that we study in this paper is a low-mass counterpart to the well studied colliding wind binaries (CWBs\footnotemark) which are systems containing two high-mass stars with strong interacting stellar winds.
Since the vast majority of high-mass stars are in binary star systems (e.g. \mbox{\citealt{2012MNRAS.424.1925C}}), CWBs are very common. 
The concept of CWBs was initially introduced by \mbox{\citet{1976SvA....20....2P}}, \mbox{\citet{1976SvAL....2..138C}}, and \mbox{\citet{1978Natur.273..645C}}.
The early studies of CWBs were dedicated mostly to trying to understand the \mbox{X-ray} emission properties of these systems, without paying attention to the detailed structure of the wind-wind collision (WWC) regions.
It was predicted that given the highly supersonic terminal wind velocities, strong shocks would be produced where the winds collide, with post-shock temperatures given by 

\footnotetext{
The term `colliding stellar winds', CSWs, is also common in the literature.
In this paper, we choose not to use either term to refer to low-mass binary systems with colliding winds. 
}

\begin{equation} \label{eqn:shocktemp}
T = \frac{3}{16} \frac{\mu m_p}{k_B} v^2,
\end{equation}

\noindent where $\mu m_p$ is the average atomic mass and $v$ is the pre-shock wind speed.
With typical wind speeds of~\mbox{$v > 1000$}~\kms, this will lead to post-shock gas being heated to temperatures of \mbox{$T > 10$~MK}, and given the high densities in the winds, significant amounts of \mbox{X-rays} will be emitted from the WWC regions.
Observational support for this idea was given by \mbox{\citet{1987ApJ...320..283P}} and \mbox{\citet{1991ApJ...368..241C}} who studied \mbox{W-R}~stars and O-type~stars respectively.
They both found higher levels of \mbox{X-ray} emission from stars in binary systems than from single stars. 
Since then, many detailed observational and theoretical studies of \mbox{X-ray} emission from WWC regions in CWBs have been carried out (e.g. \mbox{\citealt{1996MNRAS.283..589S}}; \mbox{\citealt{2009A&A...508..805G}}).

Observational detections of wind-wind interactions in CWBs in UV, visual, IR, and radio wavelengths have also been achieved (e.g. \mbox{\citealt{1985ApJ...297..255K}}; \mbox{\citealt{1993ApJ...404..281S}}; \mbox{\citealt{2012MNRAS.420.2064M}}).
Probably the most spectacular of these is the direct imaging of the pinwheel nebula coming from WR~104 (\mbox{\citealt{1999Natur.398..487T}}; \mbox{\citealt{2008ApJ...675..698T}}).
WR~104 is a \mbox{W-R} star with a close OB companion orbiting with a period of approximately 220~days. 
\mbox{\citet{2008ApJ...675..698T}} presented near-infrared Keck images at 11 epochs of an Archimedean spiral emanating from this system that can clearly be seen to spin with the orbits of the two stars.
A similar system is WR~140, which consists of a \mbox{W-R}~star and an OB star on a highly eccentric eight year orbit and shows a high level of \mbox{X-ray} emission that is clearly a result wind-wind interactions (\mbox{\citealt{1990MNRAS.243..662W}}). 
\mbox{\citet{2005ApJ...623..447D}} presented high resolution interferometric radio observations of WR~140 at 23 epochs covering almost a third of the binary orbital period.
They observed strong radio emission from the most dense part of the WWC region between the two stars that moves with their orbits.


Early physical models for the WWC~region were developed by \mbox{\citet{1982A&A...112..281H}}, \mbox{\citet{1987A&A...183..247G}}, and \mbox{\citet{1988ApJ...334.1021S}}.
Using simple ram pressure balance arguments, they were able to predict, for isotropic and steady winds, the approximate locations and shapes of the WWC regions, assuming that they are infinitely thin and neglecting Coriolis forces from the orbital motions of the two stars.
The latter assumption is only appropriate when the orbital velocities of the two stars are much smaller than the wind velocities, which can be true for large binary separations and fast winds.
In this case, the position and shape of the WWC~region is determined entirely by the ratio 

\begin{equation}
\eta \equiv \frac{\dot{M}_1 v_1 }{ \dot{M}_2 v_2},
\end{equation}

\noindent where $\dot{M}_i$ and $v_i$ are the mass loss rates and pre-shock speeds of the two winds respectively.
For values of this ratio equal to unity, e.g. for two identical winds, the WWC~region lies in the plane directly between the two stars.
As the ratio increases, the wind from the first star increasingly dominates the wind from the second star, and the WWC~region moves towards the second star and becomes increasingly bent around it. 
The WWC~region in the line joining the two stars occupies the point where the ram pressures of the two winds are equal\footnotemark. 
Specifically, the ratio of the distance between the first star and the WWC~region, $R_1$, and the distance between the second star and the WWC~region, $R_2$, is given by

\begin{equation} \label{eqn:mombalance}
\frac{R_1}{R_2} = \sqrt{\frac{\dot{M}_1 v_1}{\dot{M}_2 v_2}} = \eta^{\frac{1}{2}}.
\end{equation}

\footnotetext{
There are in fact three locations where the ram pressures could balance: the most important one is in the region where the two winds have already approximately reached terminal velocities, and the other two are in the acceleration regions of the two winds (see Fig.~2 of \mbox{\citealt{1992ApJ...386..265S}}).
The latter two are unstable, with small variations in the wind properties leading to one of the winds being completely crushed by the other.
In this situation, it might be difficult for the crushed wind to ever accelerate to high enough speeds to counter the ram pressure of the other wind.
Therefore, even for isotropic, steady winds, and circular orbits, the wind-wind interactions depend on the history of the system.
}

While instructive, the above model misses most of the physics of wind-wind interactions. 
In reality, the WWC region will not be a single infinitely thin layer, but will be a thick region of enhanced density and temperature containing a contact discontinuity and separated from the quiet winds of the two stars by two strong shock waves (\mbox{\citealt{1974Ap&SS..31..363S}}).
This is illustrated for two identical winds in Fig.~\ref{fig:CWBcartoon}.
Two important physical effects that often need to be taken into account in CWB models are cooling between the two shock waves and Coriolis forces due to the orbital motions of the stars.
In order to properly take into account all of this physics, numerical hydrodynamical models are necessary. 
The first proper hydrodynamical models of CWBs were published by \mbox{\citet{1990ApJ...362..267L}}, \mbox{\citet{1992ApJ...386..265S}}, and \mbox{\citet{1993MNRAS.260..221M}}.
These early models exploited the cylindrical symmetry that exists about the \mbox{line-of-centres} between the two stars when the orbital motions are ignored to run the simulations in 2D (alternatively, 2D simulations run in Cartesian coordinates can accurately reproduce a lot of the physics of CWBs, but have the disadvantage that the spherically expanding wind densities decrease with $r^{-1}$ instead of \mbox{with $r^{-2}$}). 
Models in 3D that were able to take into account orbital motions were first presented by \mbox{\citet{1995IAUS..163..420W}}, \mbox{\citet{1998Ap&SS.260..243W}}, and \mbox{\citet{1999IAUS..193..386P}}.
Since then, with the development of more powerful computers and more advanced numerical methods, many studies have used numerical hydrodynamic and magnetohydrodynamic models to study wind-wind interactions in high-mass star systems in increasing detail (e.g. \mbox{\citealt{2006ARep...50..722B}}; \mbox{\citealt{2007ApJ...662..582L}}; \mbox{\citealt{2009MNRAS.396.1743P}}; \mbox{\citealt{2011MNRAS.418.2618L}}; \mbox{\citealt{2014MNRAS.438.3557K}}). 

Coriolis forces break the cylindrical symmetry about the \mbox{line-of-centres} and lead to the WWC~region being wrapped around the stars and the formation of a pinwheel nebula.
When the Coriolis force is taken into account, another parameter becomes important.
This is the ratio of the orbital velocity to the wind speed, 

\begin{equation}
\zeta \equiv   \frac{   v_{\text{orb}}  }{   v_{\text{wind}}   } = \frac{1}{v_{\text{wind}}} \sqrt{\frac{G \left(M_1 + M_2 \right) }{a}},
\end{equation}

\noindent where $a$ is the orbital separation of the two stars, $v_{\text{wind}}$ is the wind speed, and $M_1$ and $M_2$ are the masses of the two stars. 
When this value is high, the shape of the wind velocity is highly influenced by Coriolis forces (\mbox{\citealt{2007ApJ...662..582L}}).
This means that orbital motions are more important for binary stars on close orbits.
For example, two 20~M$_\odot$ stars on circular orbits with a separation of 0.5~AU have orbital velocities of $\sim250$~\kms, which is a factor of a few lower than the lower-limits for typical wind speeds.
For two solar mass stars on circular orbits with a separation of 0.1~AU, the orbital velocities of the stars are $\sim$130~\kms, which is lower than typical solar wind speeds. 
For the same system, with a separation of 5~AU, the orbital velocities are $\sim$20~\kms, which is much lower than the expected wind speeds.

Another physical effect that often needs to be considered is the influence of cooling in the WWC~region, which is important for determining the separation between the two shock waves. 
For an adiabatic gas, the two shocks are relatively far apart.
The influence of cooling is to reduce the post-shock gas pressure, and therefore the WWC region thickness. 
When cooling is included, the ratio of the cooling time in the shock region to the escape time, \mbox{$\chi \equiv t_{\text{cool}} / t_{\text{esc}}$}, becomes important. 
The escape time is approximately the time that it takes for the gas to leave the WWC~region.
\mbox{\citet{1992ApJ...386..265S}} derived a simple scaling law for $\chi$.
For each wind, the importance of cooling can be characterised by

\begin{equation} \label{eqn:cooling}
\chi \sim 10^{-18} \frac{v_{\text{wind}}^4 d}{\dot{M}},
\end{equation}

\noindent where $v_{\text{wind}}$ is the wind speed in \kms, $d$ is the distance between the star and the closest shock in the \mbox{line-of-centres} between the two stars in AU, and $\dot{M}$ is the mass loss rate in \mdot.
Typically when \mbox{$\chi \ll 1$}, cooling due to radiative losses in the WWC region is important, and when \mbox{$\chi \gg 1$}, the wind in the WWC~region can be assumed to be adiabatic. 
For example, consider a binary system consisting of two stars with mass loss rates of $10^{-5}$~\mdot~and wind speeds of 1000~\kms.
For binary separations of 5~AU and 50~AU, we get values of $\chi$ of 0.025 and 2.5 respectively, corresponding approximately to radiative and adiabatic winds.
Given the same wind speeds and binary separations, for wind mass loss rates of $10^{-7}$~\mdot and $10^{-14}$~\mdot, which are typical for O-type stars and the Sun, we get values of $\chi$ of  2.5 and \mbox{$2.5\times10^7$} respectively. 
The dependence of $\chi$ on $d$ means that for binary systems on very eccentric orbits, the WWC~regions can oscillate between radiative and adiabatic states, as can be seen in the simulations of \mbox{\citet{2009MNRAS.396.1743P}}.  
Also, since this parameter can be very different for the two winds, it is possible that the wind is radiative on one side of the contact discontinuity and adiabatic on the other side (\mbox{\citealt{1992ApJ...386..265S}}). 
The reduction in the thickness of the WWC~regions due to radiative cooling has important effects on the wind dynamics.
For very thin WWC regions, non-linear thin-shell instabilities can form, leading to highly unstable and variable WWC~regions that can take on complex time-dependent geometries (\mbox{\citealt{1992ApJ...386..265S}}; \mbox{\citealt{1998MNRAS.298.1021M}}; \mbox{\citealt{2009MNRAS.396.1743P}}; \mbox{\citealt{2011MNRAS.418.2618L}}).


The development of more advanced physical models for colliding wind binaries has allowed detailed comparisons between theory and observation.
For example, \mbox{\citet{2006ARep...50..722B}} used a colliding wind model to describe the outburst of the symbiotic star Z~And as the result of the development of a wind from the white~dwarf companion that then collides with the wind from the donor star.
Of particular difficulty is explaining the observed \mbox{X-ray} properties of CWBs, which for a given massive star system, is a complex function of the orbital and wind parameters and the angle at which the system is viewed (e.g. \mbox{\citealt{1997MNRAS.292..298P}}; \mbox{\citealt{2004MNRAS.350..809S}}; \mbox{\citealt{2010MNRAS.403.1657P}}). 
In the context of \mbox{X-ray} emission, an important issue is whether or not the shocks in these systems are collisional. 
If the shocks are non-collisional, the postshock electron temperatures in the WWC~region will be lower than the ion temperatures.
If the rate at which electrons are then heated through Coulomb collisions is low, the gas might travel a significant distance before temperature equilibrium is established.  
\mbox{\citet{2000ApJ...538..808Z}} used a 2D hydrodynamical wind model to reproduce the observed \mbox{X-ray} spectra of WR~140 and found better fits when separate temperatures were assumed for the electrons and the ions in the postshock gas. 
Similar results were found for WR~140 by \mbox{\citet{2005ApJ...629..482P}} and for WR~147 by \mbox{\citet{2007MNRAS.382..886Z}}.


We have given, in this section, a brief and incomplete review of the research into colliding winds in massive binary star systems.
This is a field that has seen, and continues to see, a huge amount of activity.
A subject that has not yet been addressed in the literature is the colliding winds in low-mass star systems. 
Although there are indirect observational constraints on the winds of low-mass stars (e.g. \citealt{2000GeoRL..27..501G}; \citealt{2005ApJ...628L.143W}), due to our current inability to directly observe the winds of such stars, binary wind interactions in such systems are less interesting from an observational point of view than they are for the much more massive winds of high-mass stars.
However, they are more interesting from the perspective of star-planet interactions and planetary habitability. 
This is the subject that is discussed in the remainder of this paper.


\section{Colliding Wind Model} \label{sect:model}


In this paper, we simulate the propagation and interaction of stellar winds coming from the two components of a tight binary system composed of two solar mass stars orbiting each other on circular orbits ($e_{\text{binary}}=0$) with an orbital distance of 0.5 AU. 
For simplicity, we do not consider the effects of binary eccentricity.
The orbital period of our system is 91 days with each star travelling at approximately \mbox{60 km s$^{-1}$}. 
Classical tidal interaction theory for stars with convective envelopes (\mbox{\citealt{1975A&A....41..329Z}}; \mbox{\citealt{1977A&A....57..383Z}}) suggests that the time-scale for the synchronisation of the stellar rotation rates with the orbital period for this system is likely to be of the order of several hundred Gyrs and the timescale for circularisation is likely to be several orders of magnitude longer (meaning that our zero eccentricity system represents only one possible configuration)\footnotemark. 
Our system is therefore not tidally locked and the rotation rates for both stars will evolve independently and will therefore depend primarily on the age of the system.
We assume that each star possesses an isotropic wind with properties similar to that of the slow solar wind, which has a speed of 400~\kms \vspace{0.5mm} at 1~AU. 
This is likely to be a reasonable assumption for older (>~1~Gyr) systems where the stars have mostly spun-down to levels of magnetic activity similar to the present Sun, but is unlikely to be reasonable for young rapidly rotating stars. 

\footnotetext{
We note that an alternative mechanism for tidal synchronisation and circularisation presented by \mbox{\citet{1987ApJ...322..856T}} and \mbox{\citet{1988ApJ...324L..71T}} could lead to the synchronisation of the stellar rotation rates with the orbital period early within the lifetime of our system.
}

For our binary wind interaction simulation, we use a 3D numerical hydrodynamical code described in Section~\ref{sect:numericalmodel}.
In order for our simulation to be computationally feasible, we do not extend our computational domain to the surfaces of the two stars, but instead simulate the stars as spherical masks with radii of 0.1~AU. 
At the boundaries of these masks, we hold the wind properties constant for the entire simulation and we do not include the regions inside the masks in the simulation.
We assume that the winds accelerate to approximately their terminal velocities within these masks, and are therefore inserted into the simulations at the boundaries of these masks already fully formed.
We therefore need to estimate the properties of the slow solar wind at 0.1~AU, which we do in Section~\ref{sect:boundcon}.

\subsection{Numerical Method} \label{sect:numericalmodel}


We run our binary wind simulation using the magnetohydrodynamics code Nurgush (\mbox{\citealt{2010ARep...54.1063Z}}).
Nurgush solves the hydrodynamic equations in three dimensions, using a non-inertial coordinate system that rotates with the orbital motion of the two stars. 
We do not consider the influence of magnetic fields on the dynamics of the plasma.
Therefore, our simulation corresponds to a plasma with a very high plasma-$\beta$, where the plasma-$\beta$ is the ratio of the plasma thermal pressure to the magnetic pressure.  
Given the high plasma-$\beta$ values in the solar wind far from the Sun, this is likely to be a reasonable approximation. 
The set of equations that we solve is

\begin{equation}
\frac{\partial \rho}{\partial t} + \nabla \cdot (\rho \vec{v}) = 0,
\end{equation}

\begin{equation}
\frac{\partial \vec{v}}{\partial t} + (\vec{v} \cdot \nabla) \vec{v} = - \frac{\nabla P}{\rho} + 2 (\vec{v} \times \vec{\Omega})  - \nabla \phi,
\end{equation}

\begin{equation}
\frac{\partial \varepsilon}{\partial t} + (\vec{v} \cdot 
\nabla) \varepsilon + \frac{P}{\rho} \nabla \cdot \vec{v}
= 0,
\end{equation}

\noindent where $\rho$ is the mass density, $\vec{v}$ is the velocity, $\varepsilon$ is the energy, $P$ is the pressure, $\vec{\Omega}$ is the angular velocity of the rotating non-inertial frame of reference which we set to the orbital angular velocity of the two stars, and $\phi$ is the Roche potential. 
We solve this system of equations using the higher-order Gudonov-type differencing scheme, as described in detail by \mbox{\citet{Zhi10}} and \mbox{\citet{2010ARep...54.1063Z}}. 


We solve the equations on a 3D Cartesian grid with 512, 512, and 50 grid points in the $x$, $y$, and $z$ directions respectively.
Taking the point \mbox{$(x,y,z) = (0,0,0)$} as the centre of mass of the binary system, our computational domain extends from -5~AU to 5~AU in the $x$ and $y$ directions, and from -0.25~AU to 0.25~AU in the $z$ direction. 
The two stars are assumed to be located at \mbox{$y=0$} and \mbox{$z=0$}, and at \mbox{$x=-0.5$~AU} and \mbox{$x=0.5$~AU}. 
The stellar winds are generated by placing spherical masks of radius 0.1 AU around the two stars.
On the surfaces of these masks, we specify the densities, velocities, and temperatures of the winds, which we assume to be fully formed.
We hold these quantities constant in time by resetting them to their initial values every timestep.
For the initial conditions, we fill the rest of the computational domain with a low density and temperature gas. 
For the boundary conditions on each of the six outer boundaries, we assume outflow conditions. 
Therefore, a wind is able to flow from the two spherical masks into the computational domain, and after a short time, the initial conditions are flushed out.


In reality, by analogy with the solar wind, the expanding wind outside of the WWC~region could be heated, and inside the WWC region, both heating and cooling could be taking place. 
In simulations of winds from both single stars and binary star systems, where detailed knowledge of the required heating and radiative cooling is often not available, it is common to simulate heating and cooling by varying the value of the adiabatic index, $\gamma$. 
For a monatomic ideal gas, assuming a value of $\gamma$ of 5/3, without a heating function, corresponds to an adiabatic fluid, and assuming lower values of $\gamma$ leads to heating of the gas as it expands and cooling of the gas as it contracts.
Lower values of $\gamma$ can be used in CWB simulations to cool the post-shock plasma in WWC~region (e.g. \mbox{\citealt{1997ARep...41..794B}}).
Cooling in the WWC region would lead to a decrease in the thickness of the region between the two shock waves and the formation of thin-shell instabilities, as discussed in Section~\ref{sect:CWBintro}.  
Using Eqn.~\ref{eqn:cooling}, we estimate that the parameter $\chi$ that characterises the importance of cooling in the WWC~region is \mbox{$\sim 5 \times 10^5$} for both winds in our simulation.
We therefore do not expect radiative cooling to be significant, and set $\gamma$ to 5/3. 
This leads to a large simplification of the problem by allowing us to avoid the thin-shell instabilities that are present in highly radiative WWC~regions.

\subsection{Stellar Boundary Conditions} \label{sect:boundcon}

\begin{figure}
\includegraphics[width=0.49\textwidth]{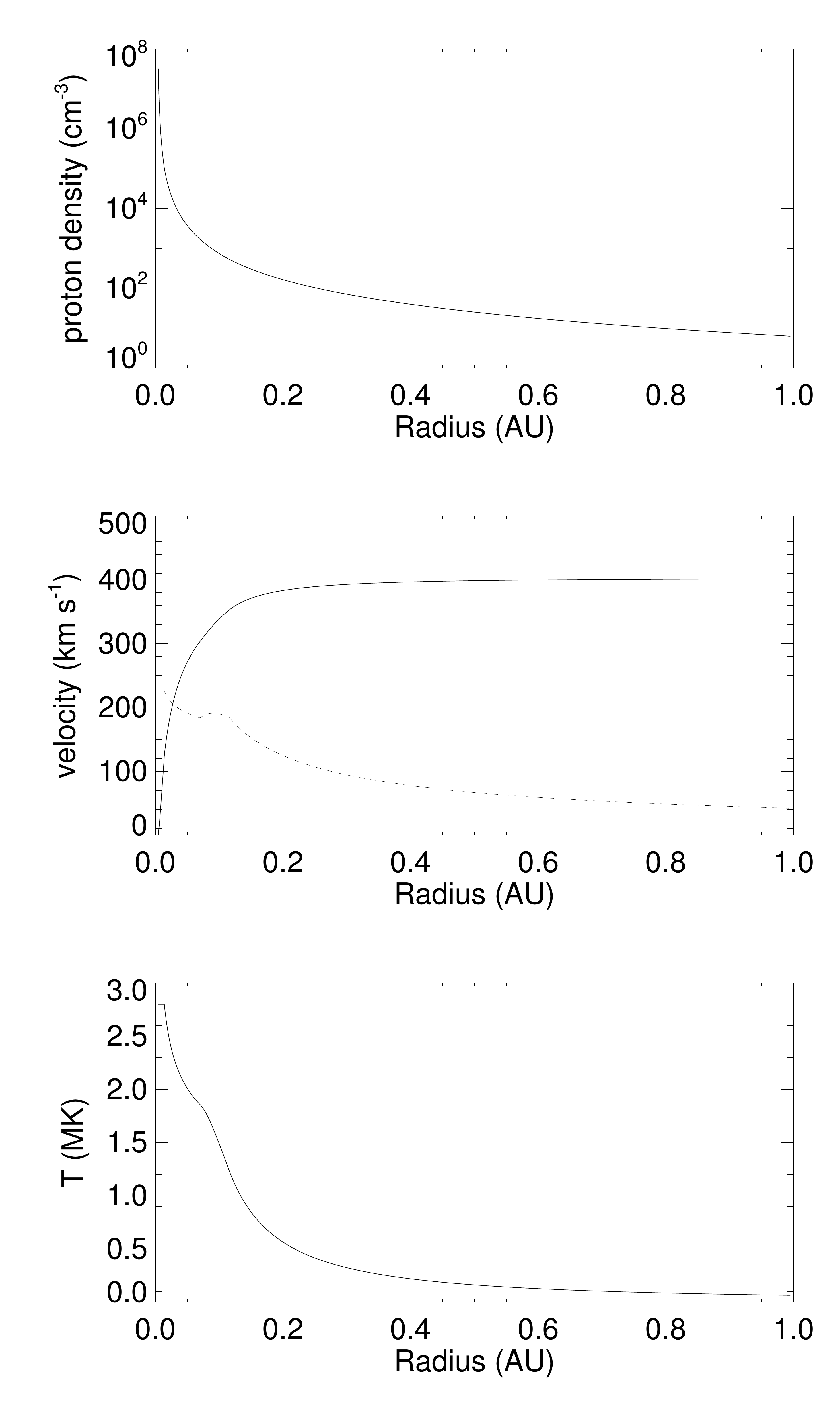}
\caption{
Plots showing our one-dimensional hydrodynamical model for the slow solar wind that we use to derive the inner boundary conditions for our binary wind simulation. 
From top to bottom, the plots show particle number density, velocity, and temperature in the wind against distance from the central star.
The dashed line in the velocity plot shows the sound speed as a function of radius.
The vertical dotted lines in each plot show the location of 0.1~AU, which is where we take the values for the inner boundary condition in the binary wind simulation. 
}
 \label{fig:slowsolarwind}
\end{figure}


In order to generate winds inside the computational domain, we need to define the wind parameters (i.e density, velocity, and temperature) at the two stellar boundaries. 
It is typical in simulations of the interactions of radiation driven winds in high-mass star systems to assume a $\beta$-law for the velocity of the wind, such that the velocity is given by \mbox{$v(r) = v_\infty (1-R_0/r)^\beta$}, where $R_0$ is some previously defined base of the wind (possibly the stellar surface), $v_\infty$ is the terminal velocity of the wind, and $\beta$ is a parameter that determines how quickly the wind accelerates to its terminal velocity (\mbox{\citealt{1986A&A...164...86P}}).
We choose instead to adopt a description of the winds from both stars that is based on the solar wind.
This is likely to be more appropriate for magnetically driven winds on low-mass stars.

To calculate the solar wind parameters at 0.1~AU, we construct a 1D hydrodynamic wind model in a spherical coordinate system using the Versatile Advection Code (VAC; \mbox{\citealt{1996ApL&C..34..245T}}; \mbox{\citealt{1999ESASP.448..389T}}).
VAC is a freely available general purpose MHD code that has been used extensively in the past for simulations of the solar wind (\mbox{\citealt{1999A&A...343..251K}}; \mbox{\citealt{2008JGRA..113.8107Z}}; \mbox{\citealt{2011AdSpR..48.1958J}}).
By only simulating the wind in 1D using an adaptive mesh that is refined close to the solar surface and becomes increasingly coarse far from the Sun, we are able to simulate the acceleration and propagation of the wind from the stellar surface to 1~AU.
We use measured properties of the slow solar wind at 1 AU to constrain the free parameters in the model.
Our model is similar to previous solar wind models in the literature (\mbox{\citealt{2011AdSpR..48.1958J}}; \mbox{\citealt{2014A&A...562A.116K}}). 

Since the fundamental driving mechanisms responsible for the heating and the acceleration of the solar wind are not known, solar wind models typically drive the wind by assuming a polytropic equation of state, \mbox{$P = K \rho^\gamma$}. 
As discussed in the previous section, setting \mbox{$\gamma < 5/3$} means that the wind is implicitly heated as it expands.
The case of \mbox{$\gamma = 1$} everywhere corresponds to an isothermal Parker wind (\mbox{\citealt{1958ApJ...128..664P}}).  
In order to accurately reproduce the solar wind, $\gamma$ must be assumed to be a function of distance from the Sun, with the wind being almost isothermal close to the Sun, and almost adiabatic further out.
Close to the solar surface, the corona and the wind are well known to be approximately isothermal, with typical values for $\gamma$ being around 1.1 (\mbox{\citealt{1988JGR....9314269S}}; \mbox{\citealt{2003ApJ...595L..57R}}).
Using Helios 1 data, \mbox{\citet{1995JGR...100...13T}} measured a value of $\gamma$ far from the Sun of 1.46. 
It would therefore make sense for us to use this value.
However, in our binary wind model, we assume a value of $\gamma$ of 5/3 in order not to add unphysical cooling in the WWC region. 
We therefore assume in our solar wind model that the wind is adiabatic far from the Sun and take \mbox{$\gamma = 1.1$} close to the Sun.
We assume that the change between these two values happens around 20~R$_\odot$, although this value is to some extent arbitrary given that we have no spacecraft measurements of the solar wind temperature structure within $\sim$0.3~AU.
In order to ensure a smooth transition between the two values of $\gamma$, we vary it from the inner value to the outer value linearly between 15~R$_\odot$ and 25~R$_\odot$.
Since we assume in our binary wind simulation that the gas is adiabatic everywhere, and set the stellar boundaries at 0.1~AU from the stellar surfaces, there is a slight discrepancy in the values of $\gamma$ between our solar wind simulation and our binary wind simulation.
However, this is only a small discrepancy that exists in a small part of the computational domain, and therefore is unlikely to significantly influence the wind dynamics.

The two free parameters in the model are then the base density of the wind, $n_0$, and the base temperature, $T_0$.
In our model, the densities of the wind at all radii, and the total mass loss rate, vary linearly with the base density. 
However, the wind velocities are only dependent on the base temperature.
The base temperature is also very important for determining the densities further from the star and therefore the mass loss rates, and the model is very sensitive to small changes in this parameter.
Since we want to reproduce the slow solar wind conditions, we take a value of $T_0$ of 2.9~MK, which leads to a wind of 402~km~s$^{-1}$ at 1~AU. 
We take a particle density at the base of the wind of $2.2 \times 10^7$~cm$^{-3}$, which gives a proton number density of 5~cm$^{-3}$ at 1~AU, consistent with typical slow wind conditions. 
Our solar wind model based on these parameters is shown in Fig.~\ref{fig:slowsolarwind}.
The wind accelerates mostly very close to the star, becomes supersonic within 5~\rsun, and by 20~R$_\odot$, is already travelling at $\sim$350~km~s$^{-1}$.
Between 20~R$_\odot$ and 1~AU, the wind accelerates much slower up to 402~km~s$^{-1}$.
Due to the way in which we vary $\gamma$ with radius, at around 20~\rsun, there is an unphysical bump in the temperature and sound speed.
This bump is small and does not influence our results.  
At 0.1~AU, this model gives a particle density of 730~cm$^{-3}$, a velocity of 340~km~s$^{-1}$, and a temperature of 1.47~MK.
We use these values as the inner boundary conditions for both stars in our binary wind simulation.


\section{Results: Wind Interactions} \label{sect:windresults}

\begin{figure*}
\includegraphics[trim=1.5cm 1cm 0.5cm 0.5cm, clip=true, width=0.49\textwidth]{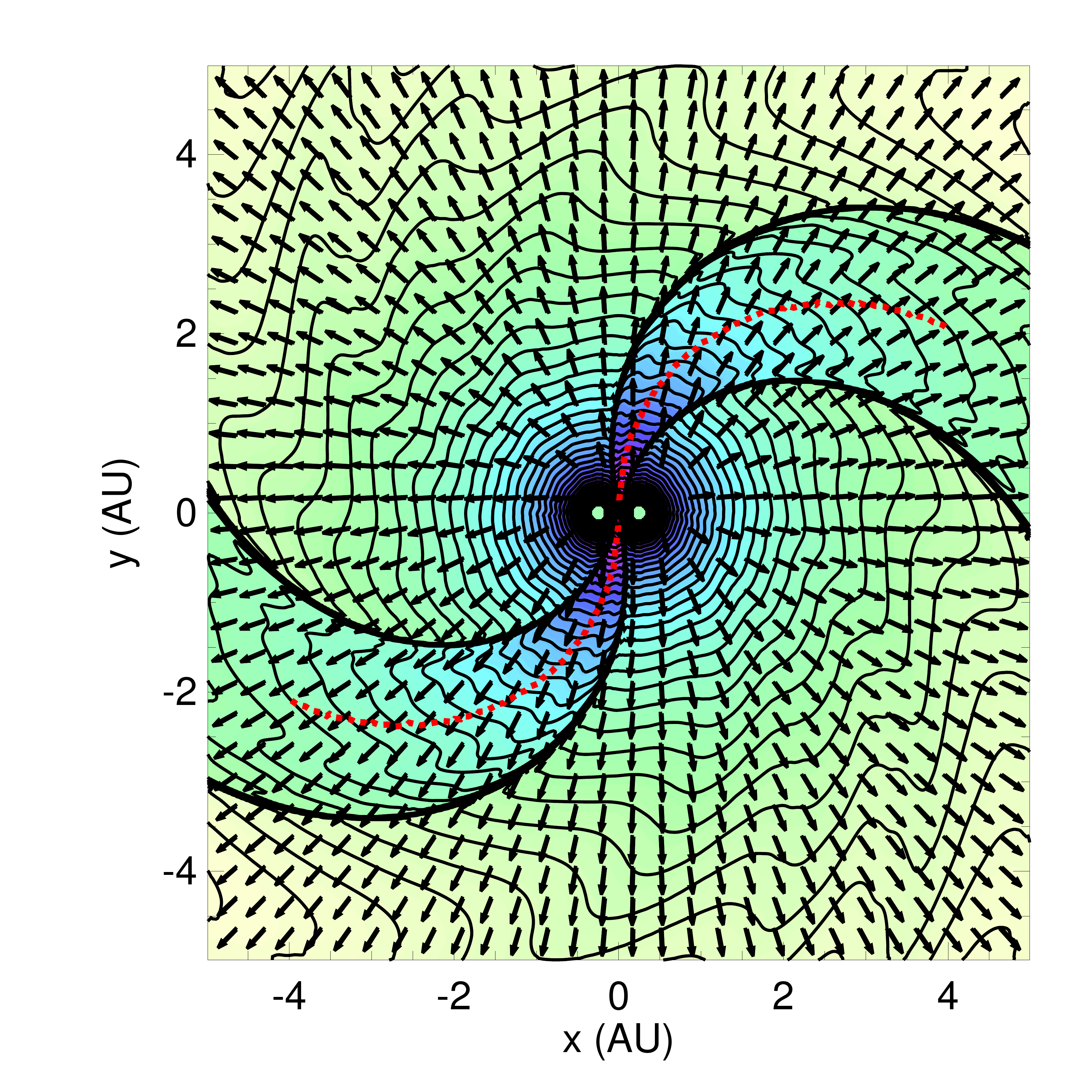}
\includegraphics[trim=1.5cm 1cm 0.5cm 0.5cm, clip=true, width=0.49\textwidth]{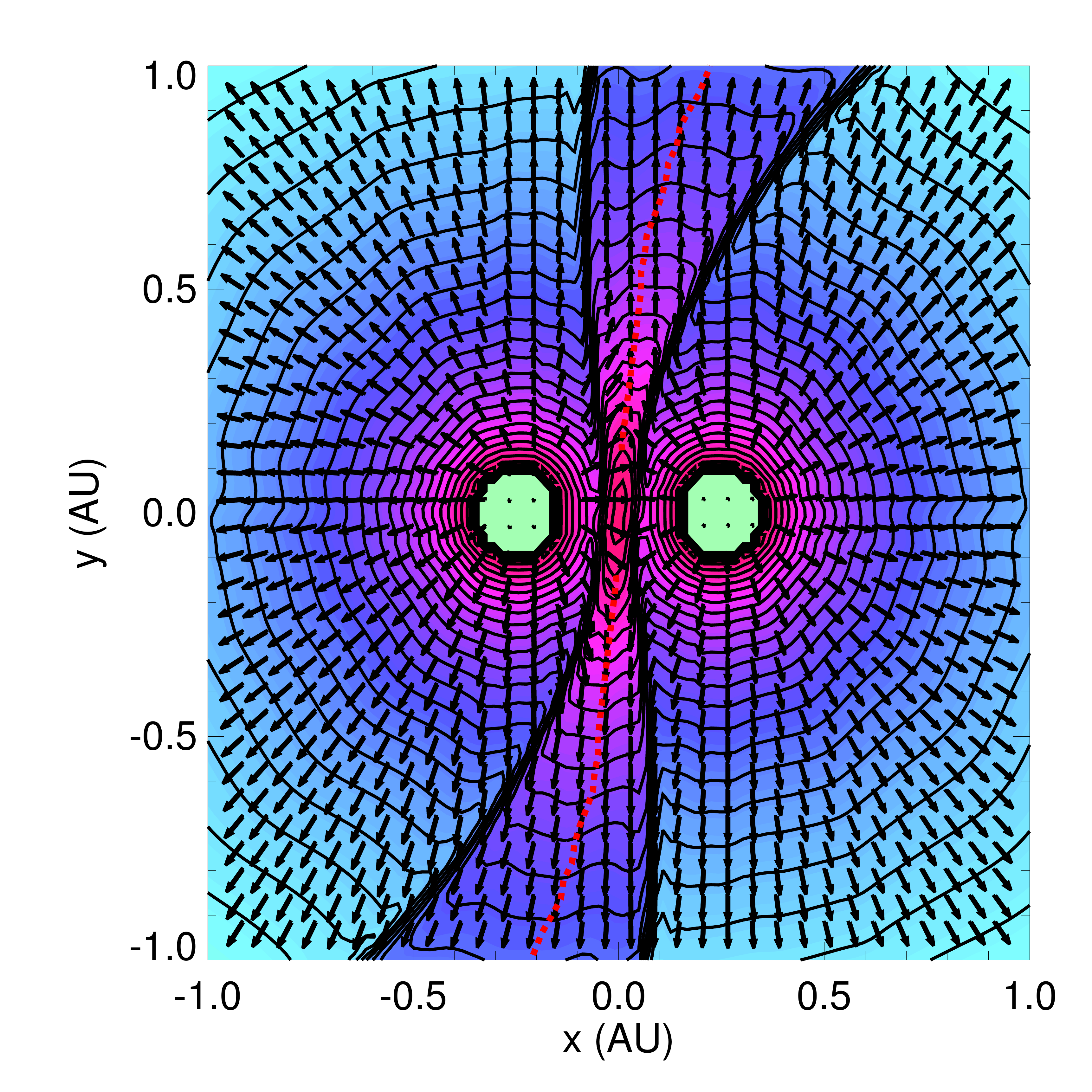}
\caption{
Cuts in the $xy$-plane showing results for the hydrodynamic binary wind simulation for the entire domain (\emph{left panel}) and for the region close to the star (\emph{right panel}).
The background colour and the contour lines show density contours and the arrows show the direction and speed of the plasma in the inertial frame of reference.
The dashed red line traces the centre of the WWC region.
}
 \label{fig:CBWcuts}
\end{figure*}


\begin{figure*}
\includegraphics[width=0.99\textwidth]{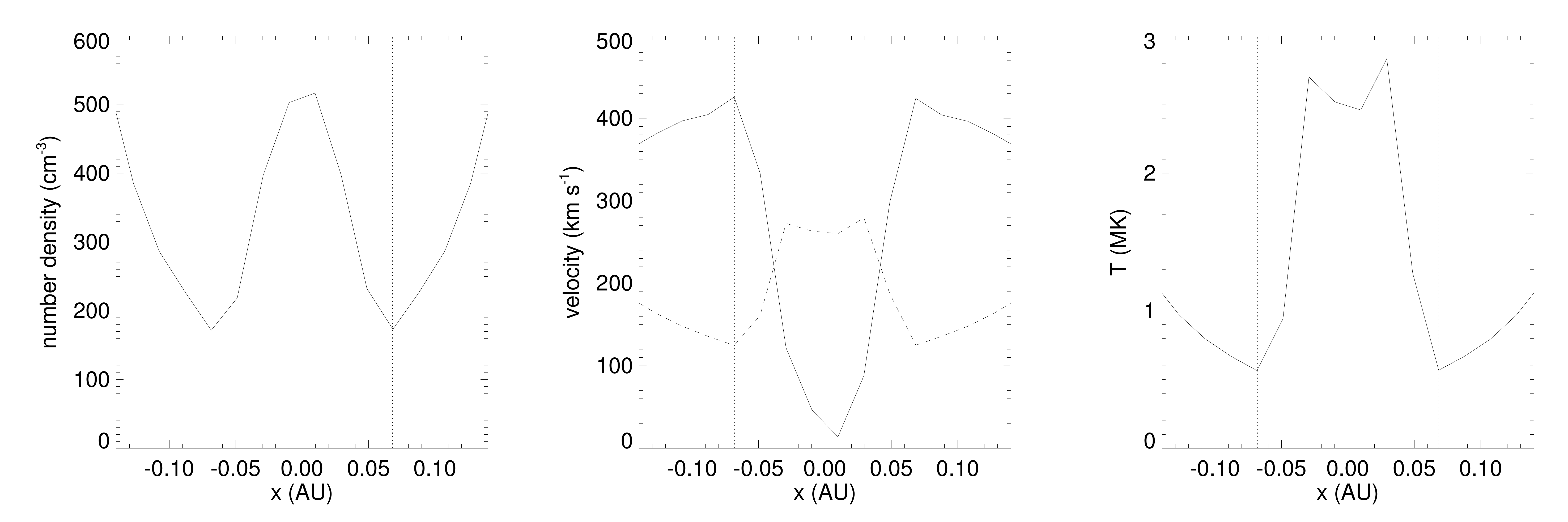}
\caption{
Plots showing particle number density (\emph{left panel}), velocity (\emph{middle panel}), and temperature (\emph{right panel}) along the \mbox{line-of-centres} directly between the two stars (see Fig~\ref{fig:CWBcartoon}).
The dashed line in the velocity plot shows the sound speed along the line. 
The vertical dotted lines represent approximately the locations of the two shocks.
}
 \label{fig:LOC}
\end{figure*}

\begin{figure*}
\includegraphics[width=0.99\textwidth]{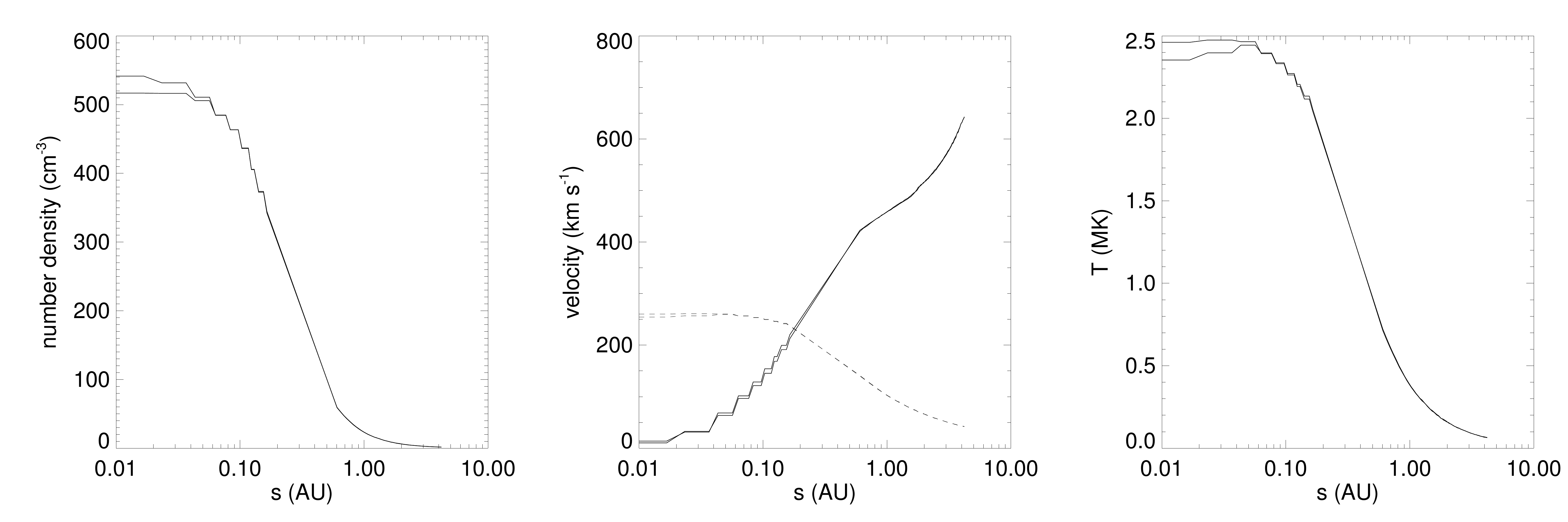}
\caption{
Plots showing particle number density (\emph{left panel}), velocity (\emph{middle panel}), and temperature (\emph{right panel}) along the line in the centre of the WWC region (i.e., the dashed red line in Fig.~\ref{fig:CBWcuts}) as a function of distance along the line.
The two lines in each plot correspond to the two directions from the centre outwards.
The dashed lines in the velocity plot show the sound speed along the lines. 
The increase in wind speed far from the centre of the system is mostly a result of Coriolis forces in the rotating frame of reference of our simulation.
}
 \label{fig:alonglines}
\end{figure*}


In Section~\ref{sect:model}, we describe our numerical model for the propagation and interaction of stellar winds from two solar mass stars with a separation of 0.5~AU. 
At the beginning of the simulation, a wind is made to propagate from the two stellar boundaries, leading to the formation of shocks.
These shocks propagate outwards, interact with each other in the region between the two stars, and propagate to the edges of the computational domain where they leave the simulation. 
After this, the simulation quickly relaxes to a steady state, and what remains is the final solution. 
It takes approximately 10,000 timesteps for the simulation to come to a steady state, which corresponds to 36~days in simulation time.
We run the simulations to 90~days and see no further change in the final result.
The time it takes for the simulation to get to a steady state is approximately 1.5 times the time for the winds emanating from the stellar boundaries to reach the outer boundaries. 
Since this initial phase has no physical significance, we do not discuss it further.


A cut through the computational domain in the orbital plane of the two stars is shown in Fig.~\ref{fig:CBWcuts}.
The background colour and the contour lines show density, and the arrows show the direction of motion of the gas. 
The form of the solution is that of a wind-wind collision (WWC) region of enhanced density and temperature, surrounded by strong shocks, as illustrated in Fig.~\ref{fig:CWBcartoon}.
Outside of the interaction region, the computational domain is filled with a quiet wind propagating outwards with an approximately constant radial velocity.
The winds from the two stars in the quiet regions are completely separated from each other by the WWC region.
The centre of the two shocks on the \mbox{line-of-centres} is directly between the two stars, as would be expected for winds with equal ram pressures.
The dashed red line in Fig.~\ref{fig:CBWcuts} traces the line directly between the two shocks. 
Due to the identical speeds of the two winds, no clear contact discontinuity can be seen. 
Across each shock, the density increases by about a factor of four, which is what we would expect for a strong adiabatic shock.
We find a density at the centre of the \mbox{line-of-centres} joining the two stars of approximately 500~cm$^{-3}$. 
As expected, the temperature also increases across each shock, leading to temperatures as high as 2.5~MK.
These temperatures are consistent with what is expected in the post-shock gas behind an adiabatic strong shock given the wind velocities (using Eqn.~\ref{eqn:shocktemp} and $v=400$~\kms\hspace{0mm} gives a temperature of approximately 2~MK).

Wind material leaving one of the stars in the direction of the other star travels almost 0.2~AU until it reaches the nearest shock.
Going through the shock, most of the kinetic energy in the wind is converted into thermal energy.
The two winds in the postshock region compress against each other and the gas quickly slows to subsonic speeds by the time it reaches the point directly between the two stars.
The wind properties along the \mbox{line-of-centres} between the two stars are shown in Fig.~\ref{fig:LOC}. 
Due to the coarseness of the grid used in the simulation, not many grid points lie along this line.
This is likely to be the reason why the profiles shown in Fig.~\ref{fig:LOC} are not completely symmetric with respect to zero, though they are almost symmetric.
A region of high pressure gas is produced which is then accelerated away from this region in the direction parallel to the two shocks. 
In Fig.~\ref{fig:alonglines}, we show the wind properties along the line directly between the two shocks in the interaction region.  
The gas is quickly accelerated by thermal pressure gradients due to the high temperatures in the post-shock gas and reaches approximately terminal velocity at around 0.5~AU from the centre of the computational domain.

The wind inside the WWC~region ends up approximately equal to the speeds of the gas outside the WWC region. 
Further out, the velocity continues to increase in the frame of reference of the two stars due to Coriolis forces. 
The similarity between the speeds of the wind inside and outside the WWC region is an inevitable consequence of our assumption that the wind is adiabatic everywhere. 
Gas streams out of the stellar boundaries with a certain amount of kinetic energy which is then converted into thermal energy as the gas passes through one of the shocks. 
The thermal energy is then converted back into kinetic energy as the gas is accelerated away from the centre of the WWC region. 
Although most of the thermal energy is reconverted into kinetic energy, the temperature in the WWC region remains a factor of a few higher than the temperature in the surrounding wind, and the velocity of the wind is always slightly lower.


The influence of the orbital motions of the stars on the wind and the WWC~region is clearly visible. 
The Coriolis force twists the WWC region into the shape of an Archimedean spiral in the $xy$-plane.  
Assuming that the $\phi$ coordinate is defined such that the line of $\phi=0$ points along the positive $x$-axis and the \mbox{$\phi = \pi/2$} line points along the positive $y$-axis, the Archimedean spiral can be described by \mbox{$r \approx 6.7 - 4.3\phi$}, where $r$ is the radial distance from the centre of the spiral, implying that the spiral makes full turns every $\sim$25~AU.
In simulations with equal winds from both stars and a circular orbit, the shape of the spiral is determined by the ratio of the wind velocity to the orbital velocity of the two stars.
The orbital velocity of the two stars is 60~\kms, giving a ratio for the speed of the wind to the orbital speed of $\sim6$.
The influence of Coriolis forces on the dynamics of colliding winds in binary systems has been studied in detail by \mbox{\citet{2007ApJ...662..582L}}.
They gave simulations with different values of the wind speed to orbital speed ratio, and our simulation shows similar results to their simulation with a similar ratio (see simulation C5 in Fig.~9 of their paper).
The Coriolis forces cause the 3D structure of the shock regions to have a more complex shape that depends on the height above the xy-plane (\mbox{\citealt{2007ApJ...662..582L}}).
Since our computational domain only extends up to 0.5~AU in the $z$-direction, such complex shapes cannot be seen.

\begin{figure}
\includegraphics[width=0.49\textwidth]{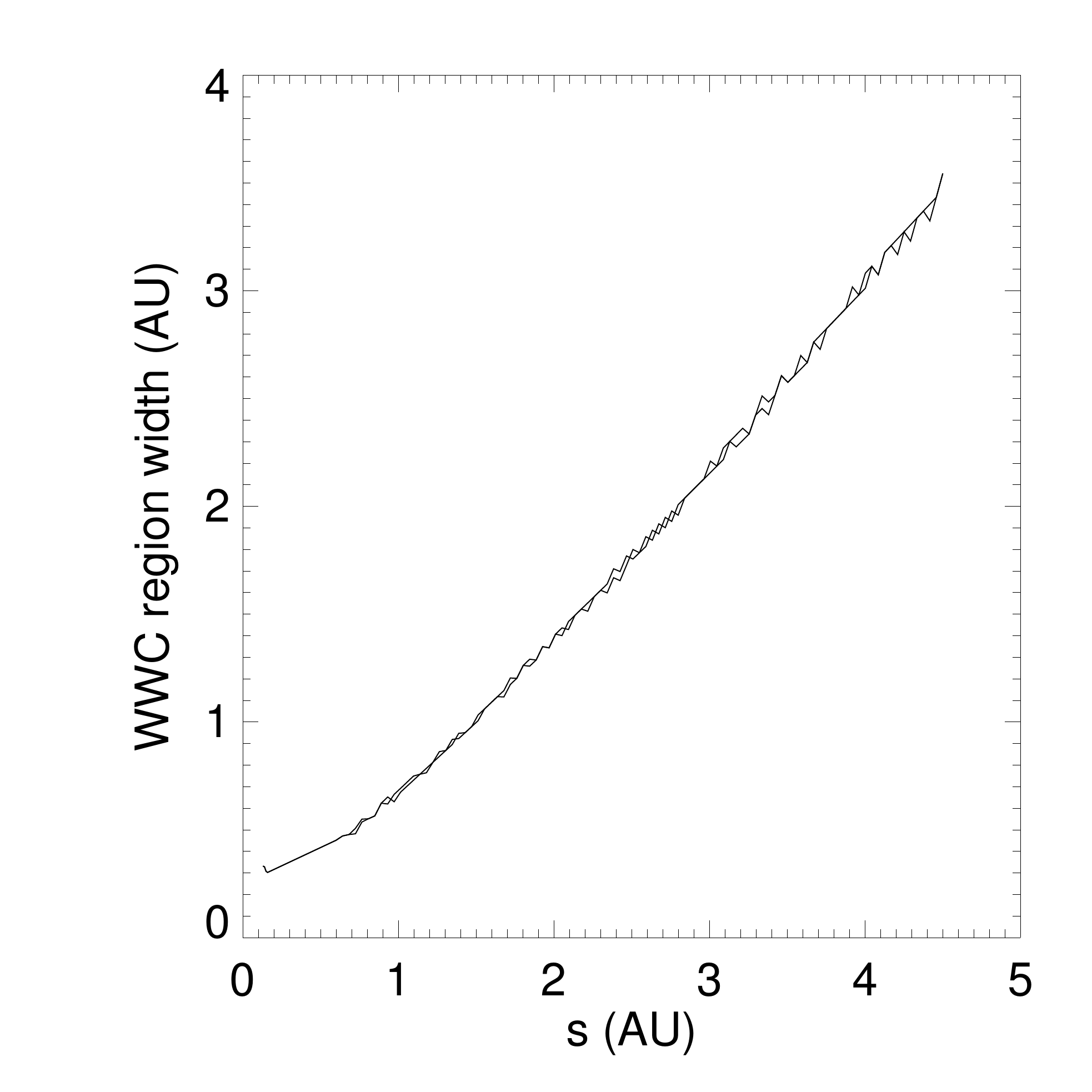}
\caption{
Plot showing the distance between the two shock waves as a function of distance along the WWC region. 
As in Fig.~\ref{fig:alonglines}, the two lines correspond to the two directions from the centre of the WWC region that we trace outwards.
}
 \label{fig:IRthickness}
\end{figure}

In Fig. \ref{fig:IRthickness}, we show how the thickness of the shock region in the orbital plane of the two stars depends on distance from the centre of the spiral. 
This is calculated by tracing the density along circles centred on the centre of the computational domain with successively expanding radii and determining along each circle the locations of the shocks. 
We then define the shock thickness as the length along this circle between the two shocks. 
The thickness of the WWC region increases with distance from the centre of the computational domain as the winds expand, with distances of several AU between the two shocks by the edge of the computational domain. 
Typically, the WWC region is an eighth of the circumference of each circle, such that a cirumbinary planet on a circular orbit will spend approximately a quarter of its orbit within the WWC region.




\section{Wind Conditions in the Habitable Zone} \label{sect:binaryhabitability}

\begin{figure}
\includegraphics[width=0.49\textwidth]{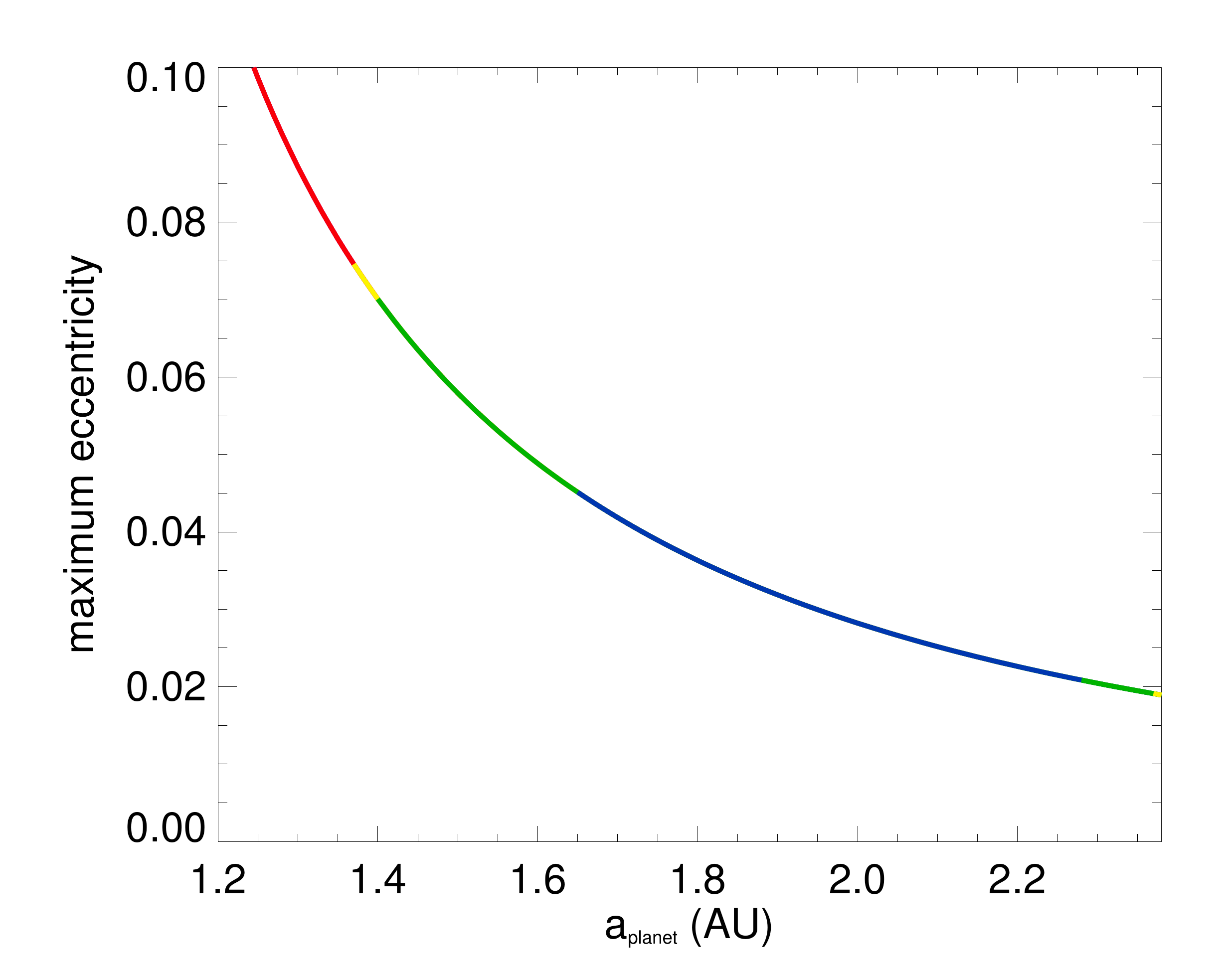}
\caption{
Plot showing the maximum orbital eccentricity of a planet as a function of semi-major axis in our binary system. 
The colours have the same meanings as in Fig.~\ref{fig:binaryhabzones1}.
Due to gravitational perturbations, potentially habitable planets around a tight binary system periodically generate and lose orbital eccentricity. 
However, in our binary system, it is unlikely that this eccentricity would have a significant influence on the wind conditions seen by a planet in the habitable zone. 
}
 \label{fig:eccentricity}
\end{figure}



In this section, we derive the location of the habitable zone in our binary system and combine this with our wind model to calculate the wind conditions a habitable circumbinary planet would see. 
In the following discussion, we assume that all planets are orbiting in the plane of the binary system and in the same direction as the two stars; we do not consider retrograde orbits. 
In our system, the two stars have a separation of 0.5~AU, and so circumstellar (\mbox{S-type}) orbits would have to be very close to one of the stars to be stable.
Such orbits are not interesting from the point of view of planetary habitability, so we therefore only consider circumbinary (\mbox{P-type}) orbits. 

The determination of the borders of the habitable zones in binary star systems has received a large amount of attention recently in the literature (e.g. \citealt{2012ApJ...752...74E}; \citealt{2012MNRAS.422.1241F}; \citealt{2013ApJ...762....7K}; \citealt{2013MNRAS.428.3104E}; \citealt{2013ApJ...777..165K}; \citealt{2013ApJ...777..166H}; \citealt{2014MNRAS.437.1352F}; \citealt{2014ApJ...780...14C}; \citealt{2014MNRAS.443..260J}), especially since several planets in such systems have been discovered by the \emph{Kepler} mission. 
Calculations for the locations of habitable zones are much more complicated when considering binary star systems than when considering single star systems.
Firstly, orbital stability must be considered; obviously a planet can only be habitable when it has a stable orbit, which in binary systems puts limits on the possible orbits that a planet can have.
When the locations of the stable orbits are known, the locations of habitable orbits, based on the classical definition of liquid water, must then be estimated. 
The calculations of the classical habitable zones are based primarily on considering the flux of stellar radiation onto the planet (i.e. the insolation), but also have some dependence on stellar spectral type. 
From planetary atmosphere models, estimates have been made for the lower and upper limits of the insolation that can lead to a planet possessing liquid water (\mbox{\citealt{1993Icar..101..108K}}; \mbox{\citealt{2013ApJ...765..131K}}). 
For a single star, these insolation limits can simply be converted into boundaries for the habitable zone, assuming approximately circular planetary orbits. 
In binary systems, the boundaries for the habitable zones are much less clear given that the stellar radiation flux that a planet receives is necessarily time variable for two reasons. 
Firstly, strong gravitational perturbations in binary systems act on a planet orbiting in the habitable zone leading to the planet periodically gaining and losing orbital eccentricity.
This eccentricity causes the planet to receive variable amounts of flux from the two stars.
Secondly, even when the planetary orbit is nearly circular, the amount of light a planet receives will vary due to changes in the spatial configurations of the two stars and the planet. 
The first effect is especially important for planets on circumstellar (\mbox{S-type}) orbits, and the latter effect is especially important for planets on circumbinary (\mbox{P-type}) orbits.

\begin{figure*}
\centering
\includegraphics[trim= 4cm 0.3cm 6mm 10mm ,clip=true, width=0.99\textwidth]{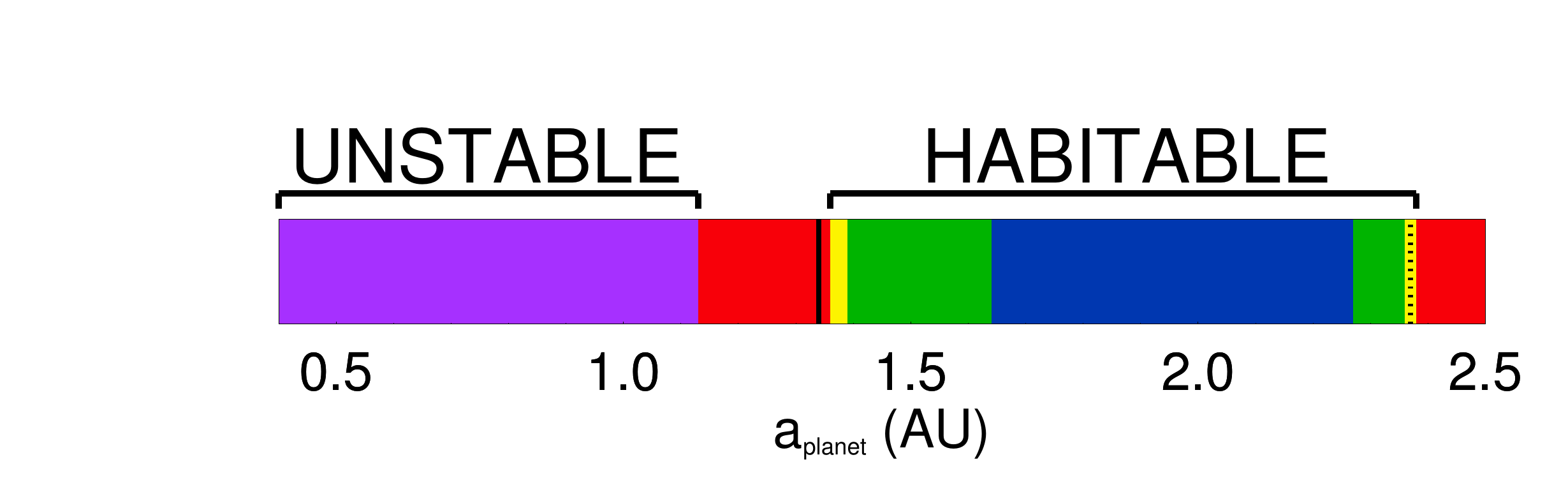}
\caption{
Figure showing the locations of the stable and habitable zones in our system, as discussed in Section~\ref{sect:binaryhabitability}.
The distances are the initial orbital distance of the planet measured from the centre-of-mass of the two stars.
The purple zone is the region in which stable planetary orbits are not possible.
Planets in the red regions are not habitable due to being either too hot or too cold for the formation of liquid water.
The blue, green, and yellow regions show the locations of the permanently habitable zones, the extended habitable zones, and the average habitable zone respectively.
The solid and dotted vertical lines show the inner and outer limits of the classical habitable zone calculated using the method of \mbox{\citet{2014ApJ...787L..29K}} for single stars, assuming a central star bolometric luminosity equal to the sums of the bolometric luminosities of the two stars in our binary system. 
}
 \label{fig:binaryhabzones1}
\end{figure*}

The first factor that we consider is orbital stability and the growth of planetary orbital eccentricity.
In a tight binary system, circumbinary (\mbox{P-type}) planetary orbits with small semi-major axes are unlikely to be stable. 
As a rule of thumb, the semi-major axis (measured from the centre of mass of the binary system) must be at least twice the distance between the two stars, which means that planetary orbits with semi-major axes below 1~AU are not possible in our system. 
Using the method described in \mbox{\citet{2002CeMDA..82..143P}}, we estimate the locations of stable orbits and the magnitude of eccentricity that a planet will gain as a function of planetary orbital distance.
For a given orbital distance, this is done by initialising a system of two solar mass stars on circular 0.5~AU orbits and a test planet and then integrating the motions of the bodies forward in time.
At any orbit, we determine orbital stability using the Fast Lyapunov Indicator (\mbox{\citealt{1997CeMDA..67...41F}}), and the eccentricity is determined directly from the simulations.  
We find that the smallest stable planetary orbit has a semi-major axis of 1.13~AU.
Planetary orbits with semi-major axes larger than 1.13~AU are likely to be stable. 
In Fig.~\ref{fig:eccentricity}, we show the maximum eccentricity that a planet will gain as a function of the semi-major axis. 
The maximum eccentricity is slightly dependent on where exactly we start the planet in our simulations.
For example, a planet started along the  $x$-axis would have a slightly lower maximum eccentricity than a planet with the same semi-major axis that is started along the $y$-axis.
In Fig.~\ref{fig:eccentricity}, we define the maximum eccentricity as the average of the values calculated from different starting positions of the planet.
Clearly, we cannot expect that the planet will be permanently on circular orbits, but will generate some eccentricity.
However, this eccentricity is not likely to be very large for a planet on a stable orbit in our system. 
For example, the centre of the habitable zone is at $\sim$2~AU where the maximum eccentricity is likely to be $\sim$0.03. 
Therefore, when considering the wind conditions that a planet will see, we make the simplifying approximation that all orbits in the habitable zone of our system are circular.

We now calculate the locations of the classical habitable zone in our system.
\mbox{\citet{1993Icar..101..108K}} defined the habitable zone as the region around a star in which planets do not exceed the following insolation limits


\begin{equation} \label{eqn:insolationlimits}
I \geq S \geq O,
\end{equation} 

\noindent where $S$ is the planetary insolation at the top of the atmosphere, and $I$ and $O$ denote upper and lower limits on the insolation that can lead to liquid water on the planet's surface.
In the case of a G2~V star, these limits are \mbox{$I=1.107$} and \mbox{$O=0.356$} in units of the solar constant (1360~W~m$^{-2}$). 
Due to temporal variations in the stellar flux incident on the planet, in some orbits, the planet will leave the habitable zone during its orbit. 
\mbox{\citet{2012ApJ...752...74E}} expanded the classical concept of habitable zones derived by \mbox{\citet{1993Icar..101..108K}}. 
Depending  on how much of the orbit is within these insolation limits, we distinguish between \emph{Permanantely Habitable Zones} (PHZ), \emph{Extended Habitable Zones} (EHZ), and \emph{Averaged Habitable Zones} (AHZ). 
The PHZ is where the planet remains within the insolation limits (i.e. satisfies Eqn.~\ref{eqn:insolationlimits}) at all times. 
The EHZ is where the planet leaves the habitable zone during its orbit, but still spends most of its orbit within the habitable zone. 
For the AHZ, we follow \mbox{\citet{2002IJAsB...1...61W}}, who used global circulation models to show that a planet leaving the habitable zone due to high orbital eccentricity might not prevent the planet from being habitable. 
Therefore, \mbox{\citet{2012ApJ...752...74E}} define the AHZ as where the time-averaged insolation, \mbox{$\langle S \rangle$}, is in the insolation limits. 
For more details, see \mbox{\citet{2012ApJ...752...74E}}.

In Fig.~\ref{fig:binaryhabzones1}, we show these different zones for our binary system, where the blue, green, and yellow areas represent the PHZ, EHZ and AHZ respectively. 
The red area shows the stable but non-habitable zone and the purple area shows the dynamically unstable region. 
The solid and dotted black lines define the inner and outer borders of the classical habitable zone for this system that we calculate using the method of \mbox{\citet{2014ApJ...787L..29K}} for a single star system with a bolometric luminosity equal to the sum of the bolometric luminosities of the two stars. 
A comparison of the two methods shows that the inner border of the classical study lies within the non-habitable zone calculated using the method of \mbox{\citet{2012ApJ...752...74E}}. 
This indicates the importance of including orbital dynamics when determining the borders of habitable zones in binary star systems.
The habitable zone shown in Fig.~\ref{fig:binaryhabzones1} defines the inner and outer parts of the AHZ and the EHZ.
The differences between these borders and the borders of the PHZ can be explained as a combination of higher eccentricities for planetary motions in these areas and the variations in the incident radiation over the planetary orbit. 
Both effects together cause parts of the orbits in these areas to evolve beyond the insolation limits.

\begin{figure}
\centering
\includegraphics[trim=1.5cm 0.5cm 1.0cm 1.0cm, clip=true, width=0.49\textwidth]{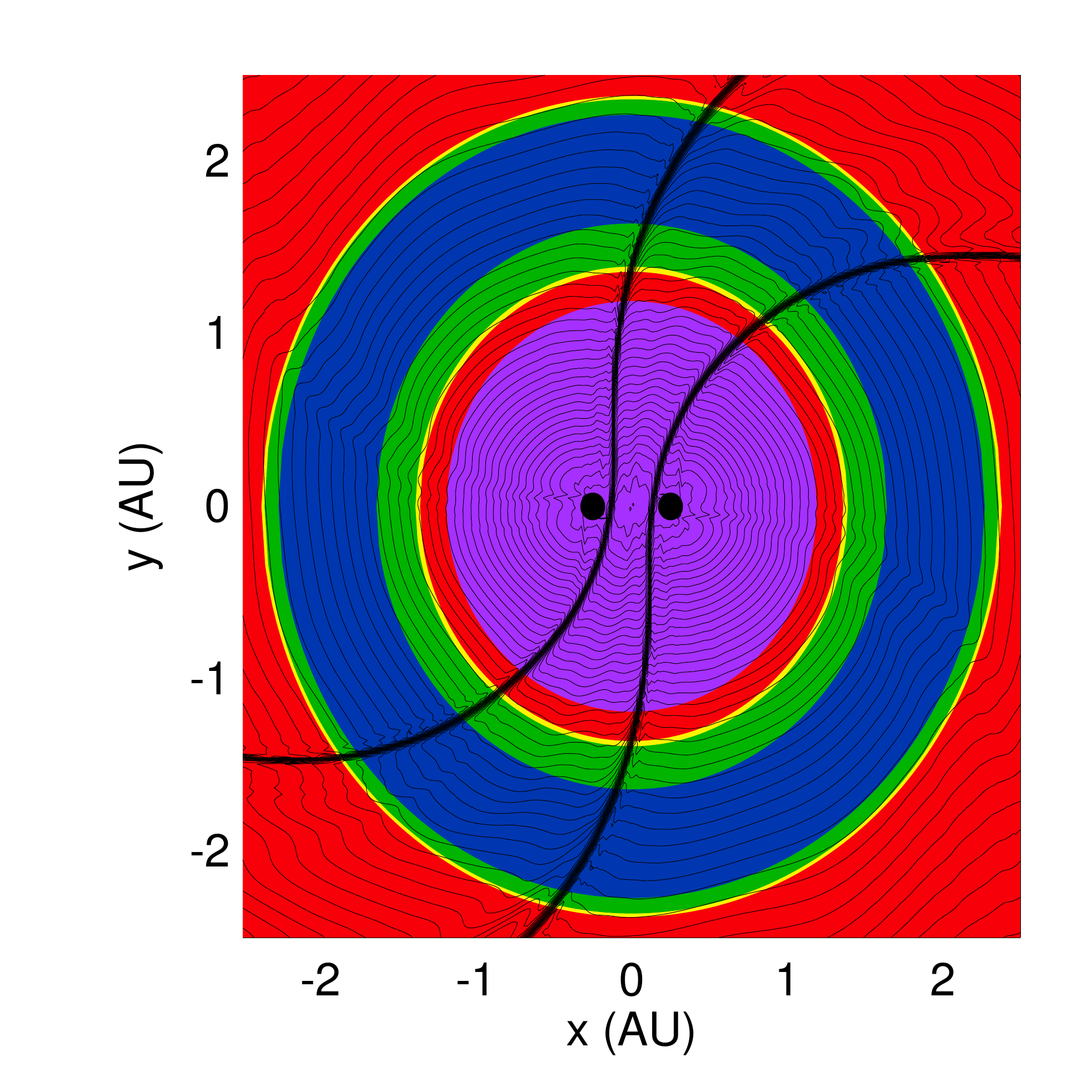}
\caption{
Figure showing the location of the habitable zone in our binary system in comparison to our wind model. 
The solid colours have the same meanings as in Fig.~\ref{fig:binaryhabzones1} and the contour lines show wind density, as in Fig.~\ref{fig:CBWcuts}.
}
 \label{fig:binaryhabzones}
\end{figure}

\begin{figure}
\includegraphics[trim=0cm 0.5cm 1.0cm 1.0cm, clip=true, width=0.49\textwidth]{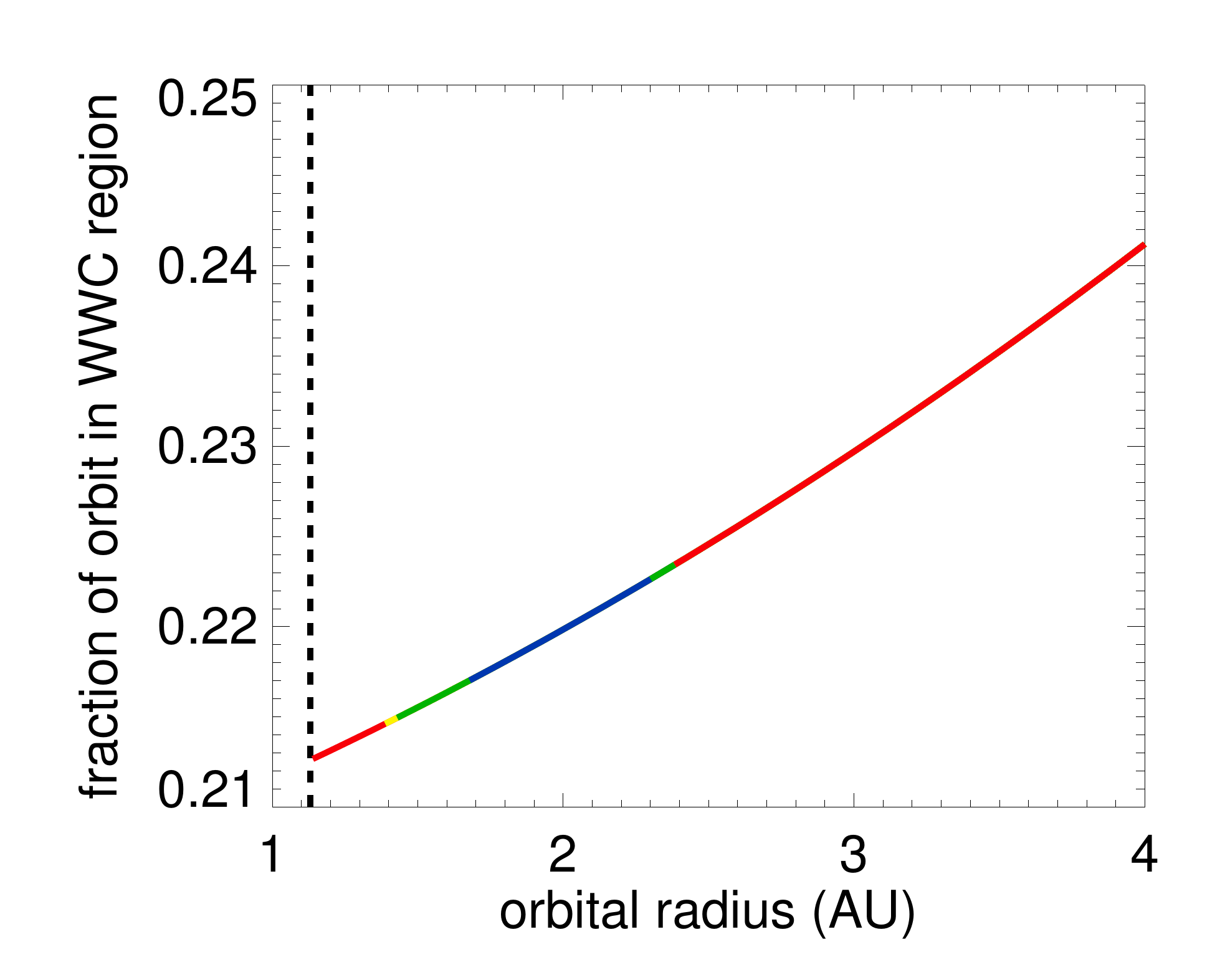}
\includegraphics[trim=0cm 0.5cm 1.0cm 1.0cm, clip=true, width=0.49\textwidth]{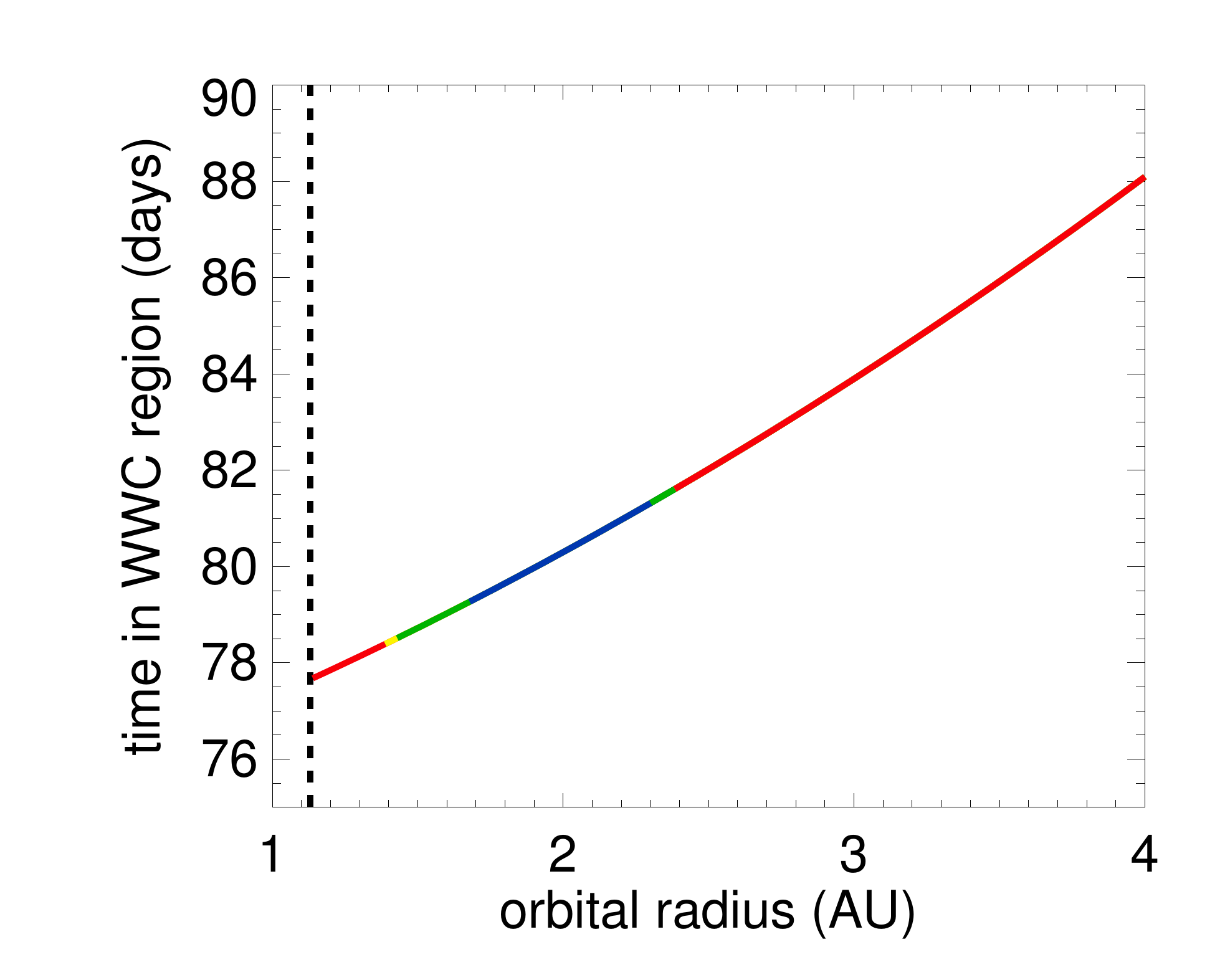}
\caption{
Plot showing the average fraction of each orbit (\mbox{\emph{upper panel}}) and the average number of days per year (\mbox{\emph{lower panel}}) that a circumbinary planet on a circular orbit would spend in the WWC region as a function of orbital radius. 
The dashed vertical lines show the minimum radius that the planet can be on a stable orbit.
The line colours correspond to the colours described in Fig.~\ref{fig:binaryhabzones1}. 
}
 \label{fig:WWCorbittime}
\end{figure}

It is important when discussing what a planet sees during an orbit to be clear how we define a planet's orbit.
When discussing our three body system, there are two frames of reference that are useful for us to consider: these are the inertial frame of reference and the frame of reference rotating about the centre of mass of the binary system with the orbit of the two stars.
In the inertial frame of reference, the two stars orbit each other with a period of 91~days (0.25~years) and a planet on a 2~AU prograde orbit, orbits with a period of 730~days (2~years).
Therefore, in this frame of reference, the planet will pass through the WWC~region fourteen times over the course of its two~year orbit.
In the rotating fame of reference, the two stars are stationary and a planet at 2~AU orbits with a period of 104~days (0.29~years) \emph{in the opposite direction} to it's motion in the inertial reference frame.
In this frame of reference, the planet passes through the WWC~region twice over the course of its orbit. 
In the following discussion, we define an `orbit' of a planet as a complete 360$^\circ$ orbit in the rotating reference frame of the binary wind simulation.
The advantage of using the rotating reference frame is that in this frame, the solution to the binary wind problem that we simulate and present in Section~\ref{sect:windresults} is not time dependent.
Interestingly, the planet passes through the WWC~region not because of its own orbital motion, but because of the much faster orbital motions of the central stars, which cause the WWC~region to catch up to and overtake the planet; in fact, the influence of the planet's own orbital motion is to decrease the frequency with which the planet passes through the WWC~region.

The main reason that we consider circumbinary orbits in this paper is that all planets on such orbits will pass through the WWC region.
In Fig.~\ref{fig:WWCorbittime}, we show the \emph{average} number of days per year (i.e. 365~days) that a circumbinary planet on a circular orbit spends within the WWC region as a function of orbital radius.
For a planet with an orbital radius of 2~AU, this is typically around 81~days, and this becomes longer for planets with larger orbital radii because of the thickening of the WWC region with distance from the two stars, as shown in Fig.~\ref{fig:IRthickness}.
At all orbital radii, at least within the computational domain, planets spend slightly less than a quarter of each orbit in the WWC region.

\begin{figure*}
\includegraphics[width=0.9\textwidth]{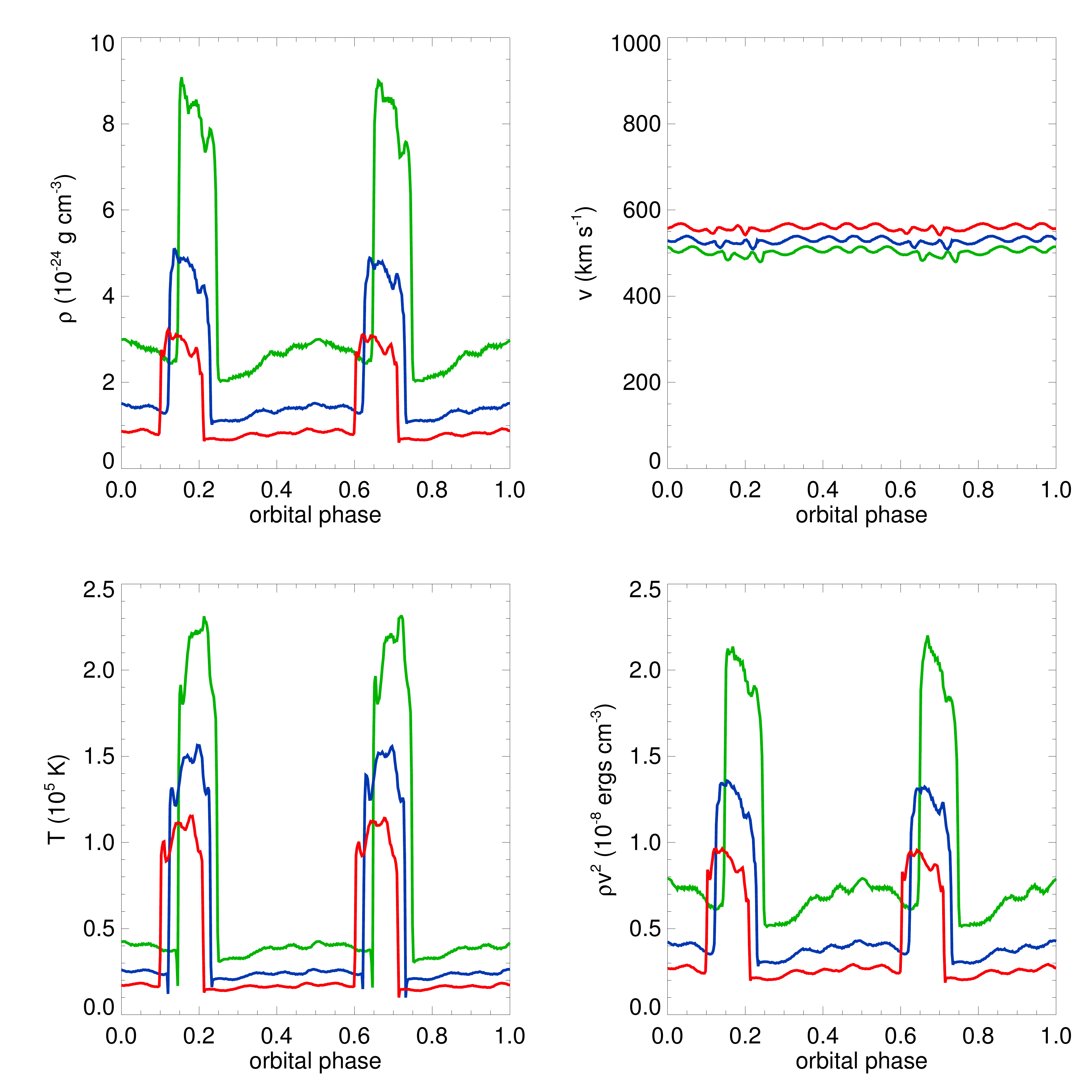}
\caption{
Variations in stellar wind parameters seen by circumbinary planets on circular orbits with orbital radii of 1.5~AU (\emph{green}), 2.0~AU (\emph{blue}), and 2.5~AU (\emph{red}).
The line colours represent which zone the planet is likely to be in and have the same meanings as in Fig.~\ref{fig:binaryhabzones1}.
}
 \label{fig:orbits}
\end{figure*}


We show in Fig.~\ref{fig:orbits} the wind properties seen by planets with orbital radii of 1.5~AU, 2.0~AU, and 2.5~AU. 
These correspond to orbits near the inner boundary, near the middle, and just outside the outer boundary of the habitable zone respectively.
At all orbital distances, as the planet passes into and out of the WWC~region, it experiences large discontinuous changes in wind density and temperature.
The wind density increases by approximately a factor of four as the planet moves into the WWC region as we expect for an adiabatic strong shock. 
A planet on a 1.5~AU orbit is exposed to wind temperatures of approximately \mbox{$4 \times 10^{4}$}~K outside of the WWC region and \mbox{$2 \times 10^5$}~K between the two shocks. 
On the other hand, the wind velocity in the WWC region is lower than outside of it, but only by a negligible amount.
The clear increase in the wind ram pressure in the WWC region is due entirely to the increase in the density. 
Outside of the WWC~region, the planet is embedded in the wind of whichever star is closer. 
Since the distance to the closest star for a planet on a circular orbit changes, the planet experiences small changes in the wind conditions even when embedded in the quiet wind. 

The elevated wind densities and temperatures in the WWC~region are likely to have some influence on the planetary upper atmosphere. 
The increase in the ram pressure will lead to a more compressed magnetosphere. 
The edge of the planetary magnetosphere pointing in the upwind direction is approximately where the ram pressure of the wind balances the magnetic pressure from the magnetosphere.
If the only parameter that changes is the wind density, $\rho_w$, then the magnetospheric radius is given by $R_m \propto \rho_w^{-1/6}$, assuming a dipole field geometry.
Therefore, a factor of four increase in the ram pressure will lead to a magnetosphere that is 80\% of its original size.
For more extended planetary atmospheres, such as those that could be present on younger planets with more massive atmospheres that are exposed to higher levels of stellar EUV emission than the current Earth, this can lead to more of the atmosphere being exposing to the stellar wind. 
At the same time, higher particle densities in the wind will increase the probability that atoms from an extended planetary atmosphere will interact with the wind, leading to the production of ions and energetic neutral atoms (ENAs) through charge exchange, and the sweeping away of planetary gas (\mbox{\citealt{2013AsBio..13.1030K}}). 
Any circumbinary planet will be exposed to these enhanced wind conditions for approximately a quarter of each orbit, and this may influence the mass loss rate and the long-term evolution of the planetary atmosphere. 
However, it is unclear whether this influence is large enough to have a significant impact on the atmosphere of a planet.


\section{Discussion} \label{sect:conclusions}


In this paper, we have studied the interactions of two identical winds in a binary star system composed of two solar mass stars on circular orbits with a binary separation of 0.5~AU. 
To do this, we have used the results of a 1D model for the slow solar wind as input into a 3D hydrodynamical wind interaction model including Coriolis forces due to the orbital motions of the two stars.
We have calculated the locations of stable and habitable zones in this system to work out what wind conditions circumbinary habitable planets see as they orbit the binary system.


In the slow solar wind model, shown in Fig.~\ref{fig:slowsolarwind}, the wind accelerates to approximately terminal velocity by 0.1~AU, meaning that in the our chosen system, the wind interaction takes place outside of the acceleration regions of the two winds.
In binary systems with much smaller separations, it is possible that this is not the case, leading to more complex wind interactions.
Our wind interaction model results in a stable steady-state solution that consists of a wind-wind collision region with enhanced density and temperature, separated from the quiet winds of the individual stars by two strong shocks. 
Given that the orbital velocities of the two stars are a significant fraction of the wind velocities, orbital motions have a significant effect on the geometries of these shocks. 
In systems with much larger separations, and therefore lower orbital velocities, the orbital motions would not be so important.


In this paper, we have chosen for simplicity to study one possible low-mass binary star system using a simplified isotropic stellar wind model.
In reality, low-mass binary systems will have ranges of orbital separations, orbital eccentricities, and combinations of stellar masses and wind properties.
At the same time, the winds themselves are likely to be more complicated than we have assumed, possessing non-isotropic and time-dependent distributions of wind properties.
At the same time, transient events, such as coronal mass ejections, are likely to additionally complicate the situation of winds in low-mass binary star systems.
These aspects will be the subject of further study.

Although we have not yet combined our results with models of planetary magnetospheres and atmospheres, it is interesting to speculate about how the interactions between the two winds in a binary system could influence a circumbinary planet.
The most interesting planets are those that orbit the binary system in the habitable zone. 
The habitable zone in our system extends from $\sim$1.4~AU to $\sim$2.4~AU from the centre of mass of the binary system. 
Inside of $\sim$1.1~AU, stable planetary orbits are not possible.
Due to gravitational perturbations, planets on stable orbits in our system would not remain on circular orbits, but would instead periodically develop and lose eccentricity, though the maximum amount of eccentricity that a planet in the habitable zone would gain is not likely to be significant for the wind properties seen by the planet. 
Combining these results with the results of the wind-wind interaction model, we are able to predict the wind conditions that a planet orbiting in the habitable zone would see.
This is shown in Fig.~\ref{fig:orbits}.
A potentially habitable planet would have to pass through discontinuous shocks four times per orbit in the rotating frame of reference of our binary wind simulations (corresponding to 28 times per orbit in the inertial frame of reference) and spend almost a quarter of each orbit in regions of enhanced wind density and temperature.


The most obvious influence of the WWC region on a planet is due to the increase in the wind ram pressures in the dense wind-wind interaction region by about a factor of four. 
This will cause the planetary magnetosphere to be compressed to $\sim$80\% of its size outside of the WWC region. 
For the current Earth, this is unlikely to lead to increased atmospheric loss due to interactions with the wind, given that the current Earth's atmosphere is a thin layer around the planetary surface, protected by a much larger magnetosphere.
However, for a much thicker hydrogen dominated atmosphere, such as the one that could have existed on the primordial Earth (\mbox{\citealt{1979E&PSL..43...22H}}; \mbox{\citealt{1980PThPh..64.1968S}}; \mbox{\citealt{2014MNRAS.439.3225L}}), this extra magnetospheric compression could lead to additional stripping of the outer layers of the planetary atmosphere. 
Therefore, it could be that in young binary systems, wind-wind interactions have a significant influence on the initial development of the a planetary atmosphere. 
This is especially true given that young stars show higher levels of magnetic activity and probably stronger winds, leading to more magnetospheric compression and therefore higher levels of atmospheric loss.

An additional influence on a planetary atmosphere by the increase in the wind density could be from an increase in the creation of energetic neutral atoms (ENAs) from charge exchange between planetary hydrogen atoms and high speed wind protons. 
\mbox{\citet{1996JGR...10126039C}} suggested that ENAs, which will not be influenced by the planetary magnetic field, could enter the atmosphere and cause additional heating as they lose their kinetic energy. 
This heating could lead to extra mass loss from the planetary atmosphere.
An additional influence of the wind-wind interactions on a planetary magnetosphere could come from the influence of the discontinuous changes in wind properties that a planet will see as it passes through the shock waves that confine the WWC region. 
An analogous situation is the influence of coronal mass ejections (CMEs) on the Earth. 
During the passage of a CME, the Earth's magnetosphere experiences reconnection events and geomagnetic storms that can be measured at the Earth's surface (\citealt{1991JGR....96.7831G}). 
One difference between these two situations is that CMEs impact the Earth from the dayside, and therefore upstream, direction, whereas the shocks at the edges of the WWC~regions in binary systems approach the planet from the direction in which the planet is moving due to it's orbital motion, which is mostly perpendicular to the upstream direction.
Whether or not this difference is significant is unclear given that in both cases, the planet sees the discontinuous changes in the wind properties from the upstream direction.

If we were to consider an Earth-mass planet with an atmosphere and magnetosphere similar to those of the current Earth in the habitable zone around a tight binary system, our results suggest that the wind-wind interactions would have little influence on the atmosphere. 
On the other hand, the influence on a much more extended hydrogen dominated proto-atmosphere around young magnetically active stars could be much larger, due to some of the processes discussed above.  
Although the situation is therefore too complicated to make definite predictions about the evolution of habitable environments in tight binary systems, we suggest that the influence of wind-wind interactions on the evolution of habitable planetary environments is a subject that deserves further attention.


\section{Acknowledgments} 

The authors thank the anonymous referee for useful comments on the paper. 
DB and AZ were supported by the Russian Foundation for Basic Research (projects 14-02-00215, 14-29-06059).
EPL, SE, CPJ, and MG acknowledge the support of the FWF NFN project S116601-N16 ``Pathways to Habitability: From Disks to Active Stars, Planets and Life'', and the related FWF NFN subprojects S116604-N16 ``Radiation \& Wind Evolution from T~Tauri Phase to ZAMS and Beyond'' and S116608-N16 ``Binary Star Systems and Habitability''. 
This publication is supported by the Austrian Science Fund (FWF).


\bibliographystyle{aa}
\bibliography{mybib}

\begin{thebibliography}{99}
\expandafter\ifx\csname natexlab\endcsname\relax\def\natexlab#1{#1}\fi

\bibitem[{{Bisikalo} {et~al.}(2006){Bisikalo}, {Boyarchuk}, {Kilpio}, {Tomov},
  \& {Tomova}}]{2006ARep...50..722B}
{Bisikalo}, D.~V., {Boyarchuk}, A.~A., {Kilpio}, E.~Y., {Tomov}, N.~A., \&
  {Tomova}, M.~T. 2006, Astronomy Reports, 50, 722

\bibitem[{{Bisikalo} {et~al.}(1997){Bisikalo}, {Boyarchuk}, {Kuznetsov}, \&
  {Chechetkin}}]{1997ARep...41..794B}
{Bisikalo}, D.~V., {Boyarchuk}, A.~A., {Kuznetsov}, O.~V., \& {Chechetkin},
  V.~M. 1997, Astronomy Reports, 41, 794

\bibitem[{{Chassefi{\`e}re}(1996)}]{1996JGR...10126039C}
{Chassefi{\`e}re}, E. 1996, \jgr, 101, 26039

\bibitem[{{Cherepashchuk}(1976)}]{1976SvAL....2..138C}
{Cherepashchuk}, A.~M. 1976, Soviet Astronomy Letters, 2, 138

\bibitem[{{Chini} {et~al.}(2012){Chini}, {Hoffmeister}, {Nasseri}, {Stahl}, \&
  {Zinnecker}}]{2012MNRAS.424.1925C}
{Chini}, R., {Hoffmeister}, V.~H., {Nasseri}, A., {Stahl}, O., \& {Zinnecker},
  H. 2012, \mnras, 424, 1925

\bibitem[{{Chlebowski} \& {Garmany}(1991)}]{1991ApJ...368..241C}
{Chlebowski}, T. \& {Garmany}, C.~D. 1991, \apj, 368, 241

\bibitem[{{Cohen}(2011)}]{2011MNRAS.417.2592C}
{Cohen}, O. 2011, \mnras, 417, 2592

\bibitem[{{Cooke} {et~al.}(1978){Cooke}, {Fabian}, \&
  {Pringle}}]{1978Natur.273..645C}
{Cooke}, B.~A., {Fabian}, A.~C., \& {Pringle}, J.~E. 1978, \nat, 273, 645

\bibitem[{{Cranmer}(2004)}]{2004ESASP.575..154C}
{Cranmer}, S.~R. 2004, in ESA Special Publication, Vol. 575, SOHO 15 Coronal
  Heating, ed. R.~W. {Walsh}, J.~{Ireland}, D.~{Danesy}, \& B.~{Fleck}, 154

\bibitem[{{Cranmer}(2009)}]{2009LRSP....6....3C}
{Cranmer}, S.~R. 2009, Living Reviews in Solar Physics, 6, 3

\bibitem[{{Cuntz}(2014)}]{2014ApJ...780...14C}
{Cuntz}, M. 2014, \apj, 780, 14

\bibitem[{{Dougherty} {et~al.}(2005){Dougherty}, {Beasley}, {Claussen},
  {Zauderer}, \& {Bolingbroke}}]{2005ApJ...623..447D}
{Dougherty}, S.~M., {Beasley}, A.~J., {Claussen}, M.~J., {Zauderer}, B.~A., \&
  {Bolingbroke}, N.~J. 2005, \apj, 623, 447

\bibitem[{{Doyle} {et~al.}(2011){Doyle}, {Carter}, {Fabrycky}, {Slawson},
  {Howell}, {Winn}, {Orosz}, {Prsa}, {Welsh}, {Quinn}, {Latham}, {Torres},
  {Buchhave}, {Marcy}, {Fortney}, {Shporer}, {Ford}, {Lissauer}, {Ragozzine},
  {Rucker}, {Batalha}, {Jenkins}, {Borucki}, {Koch}, {Middour}, {Hall},
  {McCauliff}, {Fanelli}, {Quintana}, {Holman}, {Caldwell}, {Still},
  {Stefanik}, {Brown}, {Esquerdo}, {Tang}, {Furesz}, {Geary}, {Berlind},
  {Calkins}, {Short}, {Steffen}, {Sasselov}, {Dunham}, {Cochran}, {Boss},
  {Haas}, {Buzasi}, \& {Fischer}}]{2011Sci...333.1602D}
{Doyle}, L.~R., {Carter}, J.~A., {Fabrycky}, D.~C., {et~al.} 2011, Science,
  333, 1602

\bibitem[{{Dumusque} {et~al.}(2012){Dumusque}, {Pepe}, {Lovis},
  {S{\'e}gransan}, {Sahlmann}, {Benz}, {Bouchy}, {Mayor}, {Queloz}, {Santos},
  \& {Udry}}]{2012Natur.491..207D}
{Dumusque}, X., {Pepe}, F., {Lovis}, C., {et~al.} 2012, \nat, 491, 207

\bibitem[{{Duquennoy} \& {Mayor}(1991)}]{1991A&A...248..485D}
{Duquennoy}, A. \& {Mayor}, M. 1991, \aap, 248, 485

\bibitem[{{Eggl} {et~al.}(2013){Eggl}, {Pilat-Lohinger}, {Funk},
  {Georgakarakos}, \& {Haghighipour}}]{2013MNRAS.428.3104E}
{Eggl}, S., {Pilat-Lohinger}, E., {Funk}, B., {Georgakarakos}, N., \&
  {Haghighipour}, N. 2013, \mnras, 428, 3104

\bibitem[{{Eggl} {et~al.}(2012){Eggl}, {Pilat-Lohinger}, {Georgakarakos},
  {Gyergyovits}, \& {Funk}}]{2012ApJ...752...74E}
{Eggl}, S., {Pilat-Lohinger}, E., {Georgakarakos}, N., {Gyergyovits}, M., \&
  {Funk}, B. 2012, \apj, 752, 74

\bibitem[{{Forgan}(2012)}]{2012MNRAS.422.1241F}
{Forgan}, D. 2012, \mnras, 422, 1241

\bibitem[{{Forgan}(2014)}]{2014MNRAS.437.1352F}
{Forgan}, D. 2014, \mnras, 437, 1352

\bibitem[{{Froeschl{\'e}} {et~al.}(1997){Froeschl{\'e}}, {Lega}, \&
  {Gonczi}}]{1997CeMDA..67...41F}
{Froeschl{\'e}}, C., {Lega}, E., \& {Gonczi}, R. 1997, Celestial Mechanics and
  Dynamical Astronomy, 67, 41

\bibitem[{{Gaidos} {et~al.}(2000){Gaidos}, {G{\"u}del}, \&
  {Blake}}]{2000GeoRL..27..501G}
{Gaidos}, E.~J., {G{\"u}del}, M., \& {Blake}, G.~A. 2000, \grl, 27, 501

\bibitem[{{Girard} \& {Willson}(1987)}]{1987A&A...183..247G}
{Girard}, T. \& {Willson}, L.~A. 1987, \aap, 183, 247

\bibitem[{{Gosling} {et~al.}(1991){Gosling}, {McComas}, {Phillips}, \&
  {Bame}}]{1991JGR....96.7831G}
{Gosling}, J.~T., {McComas}, D.~J., {Phillips}, J.~L., \& {Bame}, S.~J. 1991,
  \jgr, 96, 7831

\bibitem[{{Gosset} {et~al.}(2009){Gosset}, {Naz{\'e}}, {Sana}, {Rauw}, \&
  {Vreux}}]{2009A&A...508..805G}
{Gosset}, E., {Naz{\'e}}, Y., {Sana}, H., {Rauw}, G., \& {Vreux}, J.-M. 2009,
  \aap, 508, 805

\bibitem[{{Haghighipour} \& {Kaltenegger}(2013)}]{2013ApJ...777..166H}
{Haghighipour}, N. \& {Kaltenegger}, L. 2013, \apj, 777, 166

\bibitem[{{Hayashi} {et~al.}(1979){Hayashi}, {Nakazawa}, \&
  {Mizuno}}]{1979E&PSL..43...22H}
{Hayashi}, C., {Nakazawa}, K., \& {Mizuno}, H. 1979, Earth and Planetary
  Science Letters, 43, 22

\bibitem[{{Huang} \& {Weigert}(1982)}]{1982A&A...112..281H}
{Huang}, R.~Q. \& {Weigert}, A. 1982, \aap, 112, 281

\bibitem[{{Jacobs} \& {Poedts}(2011)}]{2011AdSpR..48.1958J}
{Jacobs}, C. \& {Poedts}, S. 2011, Advances in Space Research, 48, 1958

\bibitem[{{Jaime} {et~al.}(2014){Jaime}, {Aguilar}, \&
  {Pichardo}}]{2014MNRAS.443..260J}
{Jaime}, L.~G., {Aguilar}, L., \& {Pichardo}, B. 2014, \mnras, 443, 260

\bibitem[{{Kaltenegger} \& {Haghighipour}(2013)}]{2013ApJ...777..165K}
{Kaltenegger}, L. \& {Haghighipour}, N. 2013, \apj, 777, 165

\bibitem[{{Kane} \& {Hinkel}(2013)}]{2013ApJ...762....7K}
{Kane}, S.~R. \& {Hinkel}, N.~R. 2013, \apj, 762, 7

\bibitem[{{Kasting} {et~al.}(1993){Kasting}, {Whitmire}, \&
  {Reynolds}}]{1993Icar..101..108K}
{Kasting}, J.~F., {Whitmire}, D.~P., \& {Reynolds}, R.~T. 1993, \icarus, 101,
  108

\bibitem[{{Kee} {et~al.}(2014){Kee}, {Owocki}, \&
  {ud-Doula}}]{2014MNRAS.438.3557K}
{Kee}, N.~D., {Owocki}, S., \& {ud-Doula}, A. 2014, \mnras, 438, 3557

\bibitem[{{Keppens} \& {Goedbloed}(1999)}]{1999A&A...343..251K}
{Keppens}, R. \& {Goedbloed}, J.~P. 1999, \aap, 343, 251

\bibitem[{{Khodachenko} {et~al.}(2012){Khodachenko}, {Alexeev}, {Belenkaya},
  {Lammer}, {Grie{\ss}meier}, {Leitzinger}, {Odert}, {Zaqarashvili}, \&
  {Rucker}}]{2012ApJ...744...70K}
{Khodachenko}, M.~L., {Alexeev}, I., {Belenkaya}, E., {et~al.} 2012, \apj, 744,
  70

\bibitem[{{Kislyakova} {et~al.}(2014){Kislyakova}, {Johnstone}, {Odert},
  {Erkaev}, {Lammer}, {L{\"u}ftinger}, {Holmstr{\"o}m}, {Khodachenko}, \&
  {G{\"u}del}}]{2014A&A...562A.116K}
{Kislyakova}, K.~G., {Johnstone}, C.~P., {Odert}, P., {et~al.} 2014, \aap, 562,
  A116

\bibitem[{{Kislyakova} {et~al.}(2013){Kislyakova}, {Lammer}, {Holmstr{\"o}m},
  {Panchenko}, {Odert}, {Erkaev}, {Leitzinger}, {Khodachenko}, {Kulikov},
  {G{\"u}del}, \& {Hanslmeier}}]{2013AsBio..13.1030K}
{Kislyakova}, K.~G., {Lammer}, H., {Holmstr{\"o}m}, M., {et~al.} 2013,
  Astrobiology, 13, 1030

\bibitem[{{Koenigsberger} \& {Auer}(1985)}]{1985ApJ...297..255K}
{Koenigsberger}, G. \& {Auer}, L.~H. 1985, \apj, 297, 255

\bibitem[{{Kopparapu} {et~al.}(2013){Kopparapu}, {Ramirez}, {Kasting}, {Eymet},
  {Robinson}, {Mahadevan}, {Terrien}, {Domagal-Goldman}, {Meadows}, \&
  {Deshpande}}]{2013ApJ...765..131K}
{Kopparapu}, R.~K., {Ramirez}, R., {Kasting}, J.~F., {et~al.} 2013, \apj, 765,
  131

\bibitem[{{Kopparapu} {et~al.}(2014){Kopparapu}, {Ramirez}, {SchottelKotte},
  {Kasting}, {Domagal-Goldman}, \& {Eymet}}]{2014ApJ...787L..29K}
{Kopparapu}, R.~K., {Ramirez}, R.~M., {SchottelKotte}, J., {et~al.} 2014,
  \apjl, 787, L29

\bibitem[{{Kostov} {et~al.}(2014){Kostov}, {McCullough}, {Carter}, {Deleuil},
  {D{\'{\i}}az}, {Fabrycky}, {H{\'e}brard}, {Hinse}, {Mazeh}, {Orosz},
  {Tsvetanov}, \& {Welsh}}]{2014ApJ...784...14K}
{Kostov}, V.~B., {McCullough}, P.~R., {Carter}, J.~A., {et~al.} 2014, \apj,
  784, 14

\bibitem[{{Lada}(2006)}]{2006ApJ...640L..63L}
{Lada}, C.~J. 2006, \apjl, 640, L63

\bibitem[{{Lamberts} {et~al.}(2011){Lamberts}, {Fromang}, \&
  {Dubus}}]{2011MNRAS.418.2618L}
{Lamberts}, A., {Fromang}, S., \& {Dubus}, G. 2011, \mnras, 418, 2618

\bibitem[{{Lamers} \& {Cassinelli}(1999)}]{1999isw..book.....L}
{Lamers}, H.~J.~G.~L.~M. \& {Cassinelli}, J.~P. 1999, {Introduction to Stellar
  Winds}

\bibitem[{{Lammer} {et~al.}(2013){Lammer}, {Erkaev}, {Odert}, {Kislyakova},
  {Leitzinger}, \& {Khodachenko}}]{2013MNRAS.430.1247L}
{Lammer}, H., {Erkaev}, N.~V., {Odert}, P., {et~al.} 2013, \mnras, 430, 1247

\bibitem[{{Lammer} {et~al.}(2003){Lammer}, {Selsis}, {Ribas}, {Guinan},
  {Bauer}, \& {Weiss}}]{2003ApJ...598L.121L}
{Lammer}, H., {Selsis}, F., {Ribas}, I., {et~al.} 2003, \apjl, 598, L121

\bibitem[{{Lammer} {et~al.}(2014){Lammer}, {St{\"o}kl}, {Erkaev}, {Dorfi},
  {Odert}, {G{\"u}del}, {Kulikov}, {Kislyakova}, \&
  {Leitzinger}}]{2014MNRAS.439.3225L}
{Lammer}, H., {St{\"o}kl}, A., {Erkaev}, N.~V., {et~al.} 2014, \mnras, 439,
  3225

\bibitem[{{Lemaster} {et~al.}(2007){Lemaster}, {Stone}, \&
  {Gardiner}}]{2007ApJ...662..582L}
{Lemaster}, M.~N., {Stone}, J.~M., \& {Gardiner}, T.~A. 2007, \apj, 662, 582

\bibitem[{{Luo} {et~al.}(1990){Luo}, {McCray}, \& {Mac
  Low}}]{1990ApJ...362..267L}
{Luo}, D., {McCray}, R., \& {Mac Low}, M.-M. 1990, \apj, 362, 267

\bibitem[{{Madura} {et~al.}(2012){Madura}, {Gull}, {Owocki}, {Groh}, {Okazaki},
  \& {Russell}}]{2012MNRAS.420.2064M}
{Madura}, T.~I., {Gull}, T.~R., {Owocki}, S.~P., {et~al.} 2012, \mnras, 420,
  2064

\bibitem[{{Mason} {et~al.}(2013){Mason}, {Zuluaga}, {Clark}, \&
  {Cuartas-Restrepo}}]{2013ApJ...774L..26M}
{Mason}, P.~A., {Zuluaga}, J.~I., {Clark}, J.~M., \& {Cuartas-Restrepo}, P.~A.
  2013, \apjl, 774, L26

\bibitem[{{Mokiem} {et~al.}(2007){Mokiem}, {de Koter}, {Vink}, {Puls}, {Evans},
  {Smartt}, {Crowther}, {Herrero}, {Langer}, {Lennon}, {Najarro}, \&
  {Villamariz}}]{2007A&A...473..603M}
{Mokiem}, M.~R., {de Koter}, A., {Vink}, J.~S., {et~al.} 2007, \aap, 473, 603

\bibitem[{{Myasnikov} \& {Zhekov}(1993)}]{1993MNRAS.260..221M}
{Myasnikov}, A.~V. \& {Zhekov}, S.~A. 1993, \mnras, 260, 221

\bibitem[{{Myasnikov} {et~al.}(1998){Myasnikov}, {Zhekov}, \&
  {Belov}}]{1998MNRAS.298.1021M}
{Myasnikov}, A.~V., {Zhekov}, S.~A., \& {Belov}, N.~A. 1998, \mnras, 298, 1021

\bibitem[{{Parker}(1958)}]{1958ApJ...128..664P}
{Parker}, E.~N. 1958, \apj, 128, 664

\bibitem[{{Pauldrach} {et~al.}(1986){Pauldrach}, {Puls}, \&
  {Kudritzki}}]{1986A&A...164...86P}
{Pauldrach}, A., {Puls}, J., \& {Kudritzki}, R.~P. 1986, \aap, 164, 86

\bibitem[{{Pilat-Lohinger} \& {Dvorak}(2002)}]{2002CeMDA..82..143P}
{Pilat-Lohinger}, E. \& {Dvorak}, R. 2002, Celestial Mechanics and Dynamical
  Astronomy, 82, 143

\bibitem[{{Pittard}(2009)}]{2009MNRAS.396.1743P}
{Pittard}, J.~M. 2009, \mnras, 396, 1743

\bibitem[{{Pittard} \& {Parkin}(2010)}]{2010MNRAS.403.1657P}
{Pittard}, J.~M. \& {Parkin}, E.~R. 2010, \mnras, 403, 1657

\bibitem[{{Pittard} \& {Stevens}(1997)}]{1997MNRAS.292..298P}
{Pittard}, J.~M. \& {Stevens}, I.~R. 1997, \mnras, 292, 298

\bibitem[{{Pittard} \& {Stevens}(1999)}]{1999IAUS..193..386P}
{Pittard}, J.~M. \& {Stevens}, I.~R. 1999, in IAU Symposium, Vol. 193,
  Wolf-Rayet Phenomena in Massive Stars and Starburst Galaxies, ed. K.~A. {van
  der Hucht}, G.~{Koenigsberger}, \& P.~R.~J. {Eenens}, 386

\bibitem[{{Pollock}(1987)}]{1987ApJ...320..283P}
{Pollock}, A.~M.~T. 1987, \apj, 320, 283

\bibitem[{{Pollock} {et~al.}(2005){Pollock}, {Corcoran}, {Stevens}, \&
  {Williams}}]{2005ApJ...629..482P}
{Pollock}, A.~M.~T., {Corcoran}, M.~F., {Stevens}, I.~R., \& {Williams}, P.~M.
  2005, \apj, 629, 482

\bibitem[{{Prilutskii} \& {Usov}(1976)}]{1976SvA....20....2P}
{Prilutskii}, O.~F. \& {Usov}, V.~V. 1976, \sovast, 20, 2

\bibitem[{{Puls} {et~al.}(1996){Puls}, {Kudritzki}, {Herrero}, {Pauldrach},
  {Haser}, {Lennon}, {Gabler}, {Voels}, {Vilchez}, {Wachter}, \&
  {Feldmeier}}]{1996A&A...305..171P}
{Puls}, J., {Kudritzki}, R.-P., {Herrero}, A., {et~al.} 1996, \aap, 305, 171

\bibitem[{{Roussev} {et~al.}(2003){Roussev}, {Gombosi}, {Sokolov}, {Velli},
  {Manchester}, {DeZeeuw}, {Liewer}, {T{\'o}th}, \&
  {Luhmann}}]{2003ApJ...595L..57R}
{Roussev}, I.~I., {Gombosi}, T.~I., {Sokolov}, I.~V., {et~al.} 2003, \apjl,
  595, L57

\bibitem[{{Sana} {et~al.}(2004){Sana}, {Stevens}, {Gosset}, {Rauw}, \&
  {Vreux}}]{2004MNRAS.350..809S}
{Sana}, H., {Stevens}, I.~R., {Gosset}, E., {Rauw}, G., \& {Vreux}, J.-M. 2004,
  \mnras, 350, 809

\bibitem[{{Sekiya} {et~al.}(1980){Sekiya}, {Nakazawa}, \&
  {Hayashi}}]{1980PThPh..64.1968S}
{Sekiya}, M., {Nakazawa}, K., \& {Hayashi}, C. 1980, Progress of Theoretical
  Physics, 64, 1968

\bibitem[{{Shore} \& {Brown}(1988)}]{1988ApJ...334.1021S}
{Shore}, S.~N. \& {Brown}, D.~N. 1988, \apj, 334, 1021

\bibitem[{{Siscoe} \& {Heinemann}(1974)}]{1974Ap&SS..31..363S}
{Siscoe}, G.~L. \& {Heinemann}, M.~A. 1974, \apss, 31, 363

\bibitem[{{Steinolfson} \& {Hundhausen}(1988)}]{1988JGR....9314269S}
{Steinolfson}, R.~S. \& {Hundhausen}, A.~J. 1988, \jgr, 93, 14269

\bibitem[{{Stevens}(1993)}]{1993ApJ...404..281S}
{Stevens}, I.~R. 1993, \apj, 404, 281

\bibitem[{{Stevens} {et~al.}(1992){Stevens}, {Blondin}, \&
  {Pollock}}]{1992ApJ...386..265S}
{Stevens}, I.~R., {Blondin}, J.~M., \& {Pollock}, A.~M.~T. 1992, \apj, 386, 265

\bibitem[{{Stevens} {et~al.}(1996){Stevens}, {Corcoran}, {Willis}, {Skinner},
  {Pollock}, {Nagase}, \& {Koyama}}]{1996MNRAS.283..589S}
{Stevens}, I.~R., {Corcoran}, M.~F., {Willis}, A.~J., {et~al.} 1996, \mnras,
  283, 589

\bibitem[{{Tassoul}(1987)}]{1987ApJ...322..856T}
{Tassoul}, J.-L. 1987, \apj, 322, 856

\bibitem[{{Tassoul}(1988)}]{1988ApJ...324L..71T}
{Tassoul}, J.-L. 1988, \apjl, 324, L71

\bibitem[{{Tian} {et~al.}(2008){Tian}, {Kasting}, {Liu}, \&
  {Roble}}]{2008JGRE..113.5008T}
{Tian}, F., {Kasting}, J.~F., {Liu}, H.-L., \& {Roble}, R.~G. 2008, Journal of
  Geophysical Research (Planets), 113, 5008

\bibitem[{{Tian} {et~al.}(2005){Tian}, {Toon}, {Pavlov}, \& {De
  Sterck}}]{2005ApJ...621.1049T}
{Tian}, F., {Toon}, O.~B., {Pavlov}, A.~A., \& {De Sterck}, H. 2005, \apj, 621,
  1049

\bibitem[{{T{\'o}th}(1996)}]{1996ApL&C..34..245T}
{T{\'o}th}, G. 1996, Astrophysical Letters and Communications, 34, 245

\bibitem[{{T{\'o}th}(1999)}]{1999ESASP.448..389T}
{T{\'o}th}, G. 1999, in ESA Special Publication, Vol. 448, Magnetic Fields and
  Solar Processes, ed. A.~{Wilson} \& {et al.}, 389

\bibitem[{{Totten} {et~al.}(1995){Totten}, {Freeman}, \&
  {Arya}}]{1995JGR...100...13T}
{Totten}, T.~L., {Freeman}, J.~W., \& {Arya}, S. 1995, \jgr, 100, 13

\bibitem[{{Tuthill} {et~al.}(1999){Tuthill}, {Monnier}, \&
  {Danchi}}]{1999Natur.398..487T}
{Tuthill}, P.~G., {Monnier}, J.~D., \& {Danchi}, W.~C. 1999, \nat, 398, 487

\bibitem[{{Tuthill} {et~al.}(2008){Tuthill}, {Monnier}, {Lawrance}, {Danchi},
  {Owocki}, \& {Gayley}}]{2008ApJ...675..698T}
{Tuthill}, P.~G., {Monnier}, J.~D., {Lawrance}, N., {et~al.} 2008, \apj, 675,
  698

\bibitem[{{Walder}(1995)}]{1995IAUS..163..420W}
{Walder}, R. 1995, in IAU Symposium, Vol. 163, Wolf-Rayet Stars: Binaries;
  Colliding Winds; Evolution, ed. K.~A. {van der Hucht} \& P.~M. {Williams},
  420

\bibitem[{{Walder}(1998)}]{1998Ap&SS.260..243W}
{Walder}, R. 1998, \apss, 260, 243

\bibitem[{{Wang}(1995)}]{1995ApJ...449L.157W}
{Wang}, Y.-M. 1995, \apjl, 449, L157

\bibitem[{{Wang}(2010)}]{2010ApJ...715L.121W}
{Wang}, Y.-M. 2010, \apjl, 715, L121

\bibitem[{{Watson} {et~al.}(1981){Watson}, {Donahue}, \&
  {Walker}}]{1981Icar...48..150W}
{Watson}, A.~J., {Donahue}, T.~M., \& {Walker}, J.~C.~G. 1981, \icarus, 48, 150

\bibitem[{{Welsh} {et~al.}(2012){Welsh}, {Orosz}, {Carter}, {Fabrycky}, {Ford},
  {Lissauer}, {Pr{\v s}a}, {Quinn}, {Ragozzine}, {Short}, {Torres}, {Winn},
  {Doyle}, {Barclay}, {Batalha}, {Bloemen}, {Brugamyer}, {Buchhave},
  {Caldwell}, {Caldwell}, {Christiansen}, {Ciardi}, {Cochran}, {Endl},
  {Fortney}, {Gautier}, {Gilliland}, {Haas}, {Hall}, {Holman}, {Howard},
  {Howell}, {Isaacson}, {Jenkins}, {Klaus}, {Latham}, {Li}, {Marcy}, {Mazeh},
  {Quintana}, {Robertson}, {Shporer}, {Steffen}, {Windmiller}, {Koch}, \&
  {Borucki}}]{2012Natur.481..475W}
{Welsh}, W.~F., {Orosz}, J.~A., {Carter}, J.~A., {et~al.} 2012, \nat, 481, 475

\bibitem[{{Williams} \& {Pollard}(2002)}]{2002IJAsB...1...61W}
{Williams}, D.~M. \& {Pollard}, D. 2002, International Journal of Astrobiology,
  1, 61

\bibitem[{{Williams} {et~al.}(1990){Williams}, {van der Hucht}, {Pollock},
  {Florkowski}, {van der Woerd}, \& {Wamsteker}}]{1990MNRAS.243..662W}
{Williams}, P.~M., {van der Hucht}, K.~A., {Pollock}, A.~M.~T., {et~al.} 1990,
  \mnras, 243, 662

\bibitem[{{Wood} {et~al.}(2005){Wood}, {M{\"u}ller}, {Zank}, {Linsky}, \&
  {Redfield}}]{2005ApJ...628L.143W}
{Wood}, B.~E., {M{\"u}ller}, H.-R., {Zank}, G.~P., {Linsky}, J.~L., \&
  {Redfield}, S. 2005, \apjl, 628, L143

\bibitem[{{Zahn}(1975)}]{1975A&A....41..329Z}
{Zahn}, J.-P. 1975, \aap, 41, 329

\bibitem[{{Zahn}(1977)}]{1977A&A....57..383Z}
{Zahn}, J.-P. 1977, \aap, 57, 383

\bibitem[{{Zhekov}(2007)}]{2007MNRAS.382..886Z}
{Zhekov}, S.~A. 2007, \mnras, 382, 886

\bibitem[{{Zhekov} \& {Skinner}(2000)}]{2000ApJ...538..808Z}
{Zhekov}, S.~A. \& {Skinner}, S.~L. 2000, \apj, 538, 808

\bibitem[{{Zhilkin}(2010)}]{Zhi10}
{Zhilkin}, A.~G. 2010, Matem. Mod., 22, 110

\bibitem[{{Zhilkin} \& {Bisikalo}(2010)}]{2010ARep...54.1063Z}
{Zhilkin}, A.~G. \& {Bisikalo}, D.~V. 2010, Astronomy Reports, 54, 1063

\bibitem[{{Zieger} \& {Hansen}(2008)}]{2008JGRA..113.8107Z}
{Zieger}, B. \& {Hansen}, K.~C. 2008, Journal of Geophysical Research (Space
  Physics), 113, 8107

\end{thebibliography}

\end{document}